\documentclass{article}
\usepackage{arxiv}
\usepackage[utf8]{inputenc}
\usepackage[T1]{fontenc}
\usepackage[hyphens]{url}
\usepackage{graphicx}
\usepackage{natbib}
\usepackage{booktabs,array,tabularx}
\usepackage{amsmath,amssymb}
\usepackage{microtype}
\usepackage{caption,subcaption}
\usepackage{placeins}
\usepackage{hyperref}
\title{OranSim: Simulating Consumer Response\\to Social Media Campaigns Before Launch}
\date{}
\renewcommand{\shorttitle}{OranSim: Simulating Consumer Response to Social Media Campaigns Before Launch}
\hypersetup{hidelinks,
  pdftitle={OranSim: Simulating Consumer Response to Social Media Campaigns Before Launch},
  pdfkeywords={social simulation, social media marketing, agent-based modeling, Hawkes process, off-policy evaluation, uplift modeling}}
\author{%
  \textbf{Jianxiang Ma\textsuperscript{3,1}, Mingfu Zhang\textsuperscript{1,2,*}, Xiaocui Yang\textsuperscript{3}}\\
  \textbf{Yichen Gao\textsuperscript{3}, Junzhao Huang\textsuperscript{3}, Yuesong Hou\textsuperscript{3}}\\
  \textsuperscript{1}OranAI, Shenzhen 518057, China\\
  \textsuperscript{2}OranAI Ltd., City of Industry, CA 91748, USA\\
  \textsuperscript{3}School of Computer Science and Engineering,\\
  Northeastern University, Shenyang 110819, China\\
  \texttt{jianxiangma020518@gmail.com}\\
  \textsuperscript{*}Corresponding author: \texttt{cto@oran.cn}%
}
\hypersetup{pdfauthor={Jianxiang Ma, Mingfu Zhang, Xiaocui Yang, Yichen Gao, Junzhao Huang, Yuesong Hou}}

\begin{document}
\maketitle
\begin{abstract}
Social simulation studies how individual behavior and social interaction produce collective outcomes. In social media marketing, campaign actions shape which consumers encounter the content and how they respond; these responses then spread through the population. We propose \textbf{OranSim}, a social simulation framework that connects creative, creator, targeting, and budget choices to this process. Heterogeneous consumers receive exposure according to content matching and platform allocation and generate initial responses, which propagate among 60 population segments. Candidate campaigns share the initial population and aligned random numbers, making their response trajectories comparable under action changes. In a controlled synthetic campaign, doubling the budget approximately doubles reach while lowering mean content match and engagement probability among the reached consumers; mean 14-day cumulative simulated response mass rises to 1.96 times the baseline. LightGBM predictors fitted to 39,000 historical RedNote notes estimate platform engagement with log-scale $R^2$ of 0.56--0.62 in five-fold cross-validation; a separate 12,154-note corpus supplies temporal, unseen-creator, and held-out-niche test splits. Public-data experiments evaluate policy-value estimation and audience ranking, and paired synthetic outcomes test counterfactual scoring. Together, scenario trajectories and engagement estimates support campaign selection according to a prespecified marketing objective. Code is available at \url{https://github.com/OranAi-Ltd/oransim}.

\end{abstract}

\section{Introduction}
\label{sec:introduction}

Social simulation studies how the behavior of individuals, their interactions, and their environment give rise to collective outcomes. Agent-based models represent heterogeneous people and their interactions to study product diffusion, social activity, and financial markets \citep{rand2011agent,rand2021diffusion,piao2025agentsociety,yang2025twinmarket}. Social media marketing offers a concrete setting for this question: brands publish content through creators, platforms allocate exposure, and consumers' initial responses seed subsequent propagation \citep{hughes2019driving,bakshy2015exposure,bakshy2012role}. Campaign response depends on how marketing actions enter this process.

Creative choices change content matching; creator selection changes the source and audience match; targeting changes who receives exposure; and budget determines reach. These actions interact through the audience they jointly select: creative, creator, and targeting choices shape the exposure ranking, and budget determines how many of the ranked consumers receive exposure. Expanding reach therefore admits consumers with different response tendencies, so the resulting population response depends on both the number of people reached and their composition. Seeding research similarly connects firms' choice of initial recipients to participation and subsequent diffusion \citep{hinz2011seeding}. Brands choose among these options before launch, when the consumers who will be reached and their responses are still unobserved. How can a simulator carry each option through exposure, individual response, and propagation from a shared initial population?

Market-response models connect marketing inputs to aggregate outcomes \citep{hanssens2001market,jin2017bayesian}; prelaunch comparison additionally requires tracing the consumers reached by each option. Cascade models forecast propagation from post-publication events \citep{zhao2015seismic,cao2017deephawkes}; the prelaunch decision precedes those observations. Social simulators represent individual decisions and interaction, and world models compute action-conditioned environment evolution \citep{piao2025agentsociety,yang2025twinmarket,ha2018worldmodels,hafner2025mastering}. Applying this simulation perspective to marketing requires introducing creative, creator, targeting, and budget into the action space and connecting them to population dynamics.

We propose \textbf{OranSim}, a social simulation framework for social media marketing. Its central idea is to express marketing actions as changes to content matching, exposure allocation, and reach, then propagate the initial responses of the selected audience. Consumer interests, activity, personality, and response states determine exposure and behavioral probabilities, and Figure~\ref{fig:overview} marks where each attribute enters. Initial responses propagate between demographic segments. Candidate scenarios share their initial population, mechanism parameters, and individual random numbers, so action changes can be compared on the same people.

Campaign objectives are stated in platform engagement counts: reads, likes, collects (saves), and comments. Predictors learned from historical RedNote posts, called notes, estimate these counts from the note content, creator, and timing of each scenario; the objective specifies how to use them alongside the scenario's simulated trajectory. In the controlled case, a collect-first objective ranks options by predicted collects and orders options with equal predictions by cumulative simulated response. This use of historical observations alongside simulation follows a prelaunch forecasting tradition in marketing \citep{trusov2013prelaunch}.

Our contributions are:
\begin{enumerate}
\item A marketing social simulation that maps campaign actions to the exposure and response of heterogeneous consumers, and connects their initial responses to population dynamics through intergroup propagation.
\item Controlled scenario comparison from shared initial conditions. A skincare campaign traces how doubling the budget expands reach, changes audience composition, and raises mean cumulative simulated response mass to 1.96 times the baseline, then applies a prespecified objective to select a campaign.
\item Evaluation of the predictors and estimators used in campaign comparison: platform engagement prediction on private RedNote notes, policy-value estimation and audience ranking on public data, and counterfactual scoring on paired synthetic outcomes, with fixed scenarios and a deterministic output snapshot for reproduction.
\end{enumerate}

\section{Related Work}
\label{sec:related}

\paragraph{Social simulation.}
Generative Agents, RecAgent, OASIS, and AgentSociety model individual behavior and interaction in social and recommendation environments \citep{park2023generative,wang2023recagent,yang2024oasis,piao2025agentsociety}. TwinMarket connects investor decisions, information propagation, and collective market dynamics \citep{yang2025twinmarket}. Language-model studies examine behavioral responses and human alignment \citep{argyle2023out,aher2023using,brand2024llm,filippas2024simulated,ye2026stop,hu2026simbench}; world models, recommendation simulators, and population frameworks support action-conditioned prediction at different scales \citep{ha2018worldmodels,hafner2025mastering,zhao2023kuaisim,zhang2024generative,zhang2025llmsimulator,zhang2025socioverse,gao2024survey}. OranSim represents a campaign as an action whose consequences unfold through statistical individual responses and intergroup propagation.

\paragraph{Consumer interaction and market diffusion.}
The Bass model describes new-product adoption through innovation and imitation \citep{bass1969new}, and agent-based marketing links consumer heterogeneity and local interaction to aggregate diffusion \citep{rand2011agent,rand2021diffusion}. Segmentation, persuasion, and interpersonal influence inform these relationships \citep{mccarthy1960basic,kotler2016marketing,petty1986communication,katz1955personal,krugman1972why}. Seeding strategies connect the choice of initial recipients to participation and reach \citep{hinz2011seeding}. For prelaunch forecasting, pretest-market models estimate new-product sales from consumer responses measured before launch \citep{silk1978pretest}, and synthetic-network priors combine simulation with historical diffusion \citep{trusov2013prelaunch}. OranSim uses campaign choices to determine the initial audience and its response, then computes propagation over that population.

\paragraph{Outcome prediction and decision evaluation.}
Market-response models estimate aggregate outcomes from marketing inputs \citep{hanssens2001market,jin2017bayesian}. Hawkes processes and cascade predictors model propagation from observed events \citep{hawkes1971spectra,zhao2015seismic,li2017deepcas,cao2017deephawkes,wang2017cascade}. Off-policy estimators use logged assignments to estimate policy value \citep{li2011unbiased,dudik2011doubly,swaminathan2015self,wang2017optimal}, with KuaiRand and Open Bandit providing public evaluation data \citep{gao2022kuairand,saito2020open}. Treatment-effect learners rank audiences by incremental response \citep{kunzel2019metalearners,nie2021quasi,athey2019generalized}. Our evaluation uses these outcome and decision measures.

\section{OranSim}
\label{sec:method}

\begin{figure}[t]
\centering
\includegraphics[width=\linewidth]{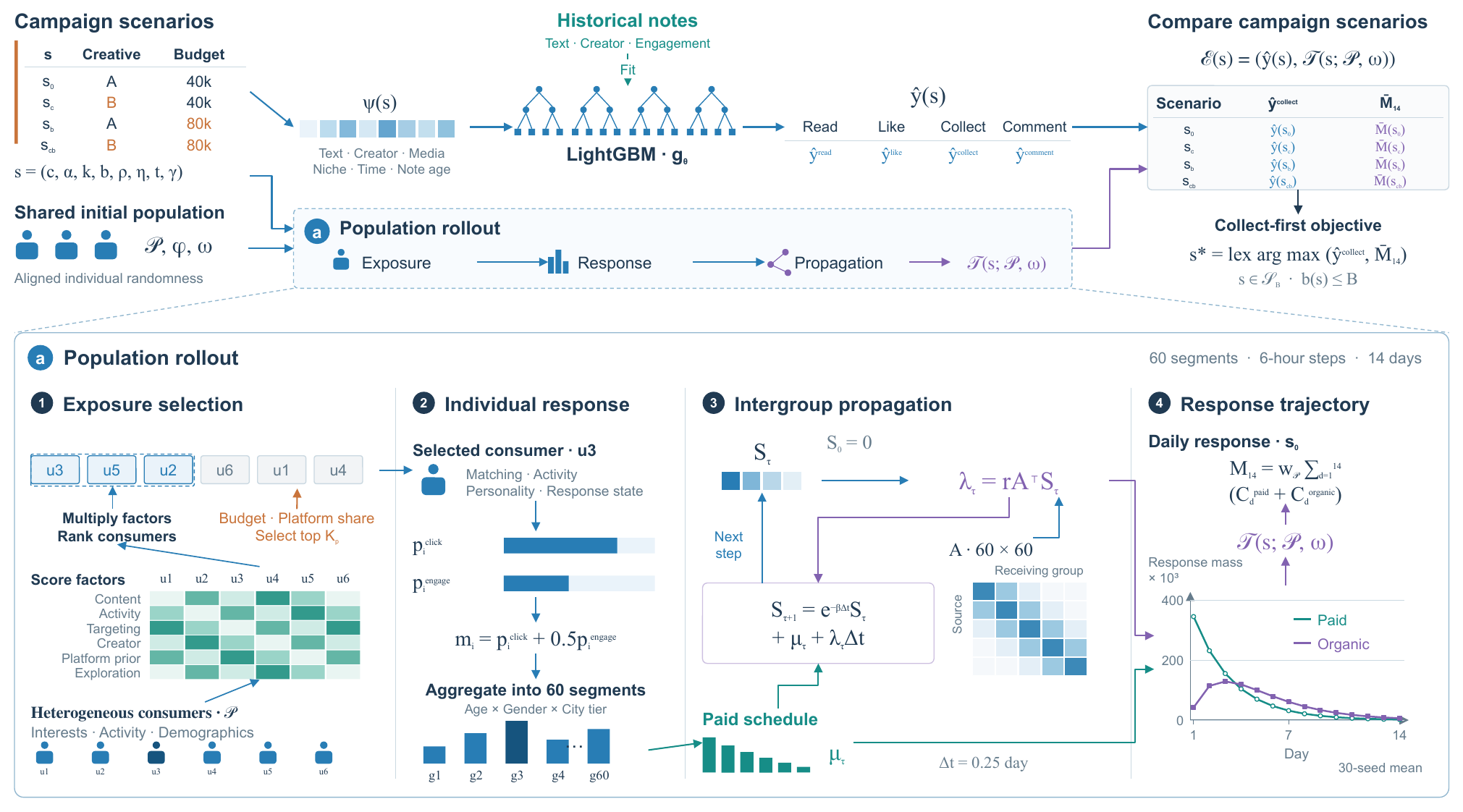}
\caption{OranSim overview: note-engagement prediction and population rollout for each scenario (top), the four rollout stages (bottom). Curves are 30-seed means for baseline $s_0$ (Section~\ref{sec:case}).}
\label{fig:overview}
\end{figure}

OranSim computes how a campaign changes the response of a simulated society. A population of heterogeneous consumers supplies interests, activity, and behavioral attributes; exposure allocation selects the initial audience; individual responses seed propagation between demographic segments. Historical-note predictors estimate platform engagement for the same scenario, and the campaign objective specifies how to use the engagement estimates and rollout trajectory. Figure~\ref{fig:overview} shows the computation and the consumer attributes used at each stage.

\subsection{Simulated Society for Marketing}
\label{subsec:society}

The population $\mathcal P$ contains consumers with demographic attributes, interest vectors, Big Five personality traits, platform activity, and response states. These attributes enter distinct parts of the simulation. Interests determine content and creator matching; demographics determine targeting weights and platform audience priors; platform activity affects exposure ranking and response probability. Openness and neuroticism enter the statistical response model, and a response-state coordinate supplies a fatigue proxy. Thus two consumers who receive the same creative can have different response probabilities. Appendix~\ref{app:population} specifies population generation.

Age band, gender, and city tier also define $6\times2\times5=60$ segments. A matrix $A$ describes the influence of each source segment on receiving segments, assigning greater weight within a segment and between demographically similar segments. Individual responses are aggregated into these segments before propagation.

\subsection{From Individual Exposure to Population Response}
\label{subsec:mechanisms}

For consumer $i$ on platform $p$, the exposure score multiplies content matching, platform activity, targeting weight, creator matching, the platform audience prior, and a seeded exploration perturbation. Consumers are selected in decreasing score order. With budget $b$, platform share $\rho_p$, population scaling weight $w_{\mathcal P}$, and cost per thousand impressions $\mathrm{CPM}_p$, the number selected is
\begin{equation}
K_p=\min\!\left(|\mathcal P|,\left\lfloor\frac{1000b\rho_p}{w_{\mathcal P}\mathrm{CPM}_p}\right\rfloor\right).
\label{eq:reach}
\end{equation}
Here $w_{\mathcal P}$ is the ratio of represented to simulated population size. Increasing the budget extends selection further down the ranking, changing both reach and the composition of the exposed audience. Appendix~\ref{app:exposure} defines each factor, including the hashed title representation for content matching.

For each selected consumer, a logistic behavior model computes click probability $p_i^{\mathrm{click}}$ from matching, activity, targeting, creator, personality, and state features. Multiplying this probability by a logistic post-click engagement propensity gives $p_i^{\mathrm{engage}}$. The initial response mass $p_i^{\mathrm{click}}+0.5p_i^{\mathrm{engage}}$ counts expected clicks and weights expected engagement by one half. We sum these masses within each segment and distribute them over time according to the paid schedule, obtaining the segment vector $\boldsymbol\mu_\tau$ at step $\tau$. The model coefficients and schedule are specified in Appendix~\ref{app:response}.

A segmented Hawkes process \citep{hawkes1971spectra} propagates these initial responses. Let $\mathbf S_\tau$ be accumulated excitation at the start of a step, and let $A_{ij}$ be the influence of source segment $i$ on segment $j$. The organic response rate and next state are
\begin{equation}
\boldsymbol\lambda_\tau=rA^\top\mathbf S_\tau,\qquad
\mathbf S_{\tau+1}=e^{-\beta\Delta t}\mathbf S_\tau+\boldsymbol\mu_\tau+\boldsymbol\lambda_\tau\Delta t,
\quad \mathbf S_0=\mathbf 0,
\label{eq:dynamics}
\end{equation}
where $r$ is the branching parameter that scales propagation excitation, $\beta$ is the decay rate, and $\Delta t$ is measured in days. Each step combines retained excitation, current paid input, and organic response induced by earlier activity. Summing $\boldsymbol\mu_\tau$ and $\boldsymbol\lambda_\tau\Delta t$ over segments and within each day gives daily paid and organic response masses $C_d^{\mathrm{paid}}$ and $C_d^{\mathrm{organic}}$. Aggregating the steps within each day separately for each segment gives its response trajectory.

\subsection{Scenario Rollouts of Marketing Actions}
\label{subsec:scenarios}

A campaign scenario is
\begin{equation}
s=(c,\alpha,k,b,\boldsymbol\rho,\eta,t,\gamma),
\label{eq:scenario}
\end{equation}
where $c$ contains the title, body, media, niche, and hashtags; $\alpha$ specifies the audience; $k$ describes the creator; $b$ is the budget; and $\boldsymbol\rho$ contains nonnegative platform budget shares summing to one. The exposure configuration $\eta$ gives platform priors, exploration strength, costs, and the paid schedule. Publication time is $t$; $\gamma\in[0,180]$ is the note age in days at which outcomes are predicted.

Each action has a concrete point of entry. Creative $c$ changes content matching in exposure and response; creator $k$ supplies niche and audience-interest matching. Audience setting $\alpha$ changes demographic targeting weights. Budget $b$ and shares $\boldsymbol\rho$ determine reach through Equation~\eqref{eq:reach}. For example, replacing a skincare creative recomputes content matching, the selected audience, and its responses while holding the budget and initial population fixed.

The rollout $\mathcal T(s;\mathcal P,\omega)=R_\phi(s,\mathcal P;\omega)$ performs exposure, response, and propagation in sequence, where $\phi$ contains their mechanism parameters and $\omega$ is a random seed. Candidate scenarios share $\mathcal P$ and $\phi$, and each individual uses aligned random numbers across scenarios. Exposure and propagation states are computed separately for each candidate. This pairing reduces variation due to population resampling when comparing action changes.

The cumulative simulated response over $H$ days is
\begin{equation}
M_H(s,\omega)=w_{\mathcal P}\sum_{d=1}^{H}\bigl(C_d^{\mathrm{paid}}(s,\omega)+C_d^{\mathrm{organic}}(s,\omega)\bigr).
\label{eq:mass}
\end{equation}
It measures weighted response mass in the represented population. For baseline $s_0$ and candidate $s_1$, the controlled difference is
\begin{equation}
\Delta_{\mathrm{sim}}(s_1,s_0;\omega)=M_{14}(s_1,\omega)-M_{14}(s_0,\omega).
\label{eq:contrast}
\end{equation}
We average this difference over common seeds. Daily trajectories retain the timing of response, while $M_{14}$ compares the total over the same window.

\subsection{Observed Engagement Prediction}
\label{subsec:outcomes}

Historical notes pair the content, creator, and timing of published notes with their engagement counts. For each note, the predictor concatenates principal components of title and body representations with log follower count, duration, encoded media type and niche, hashtag features, cyclic publication-time features, and normalized note age $\gamma/180$; for training notes, age counts the days from publication to the fitting date, capped at 180. Denote this scenario-to-feature map by $\psi(s)$. We compare configurations without and with note age, called v3 and v3.1-pg, respectively. Each trains four separate LightGBM regressors \citep{ke2017lightgbm} on $\log(1+y)$, one each for reads, likes, collects, and comments, and reports $\exp(\hat z)-1$ as the predicted count for each log-scale prediction $\hat z$. Separate outcomes allow a campaign to prioritize its chosen engagement objective.

\subsection{Prelaunch Comparison}
\label{subsec:comparison}

For a scenario, the evaluation record combines its engagement predictions and trajectory:
\begin{equation}
\mathcal E(s;\mathcal P,\omega)=\bigl(\hat{\mathbf y}(s),\mathcal T(s;\mathcal P,\omega)\bigr)
=\bigl(g_\theta(\psi(s)),R_\phi(s,\mathcal P;\omega)\bigr),
\label{eq:joint}
\end{equation}
where $g_\theta$ denotes the four fitted regressors. Both computations receive the same scenario and describe different objects: $\hat{\mathbf y}(s)$ predicts the engagement of the campaign note, and the rollout traces the consumers the campaign reaches and their responses. Audience, budget, platform shares, and exposure configuration enter only the rollout, where they shape who is reached; a budget-only change therefore leaves $\psi(s)$ unchanged and selects more consumers. The campaign objective determines how to compare the resulting quantities.

For the collect-first objective with budget cap $B$, let $\mathcal S_B=\{s\in\mathcal S:b(s)\le B\}$. We select
\begin{equation}
s^\star=\underset{s\in\mathcal S_B}{\operatorname{lex\,arg\,max}}
\bigl(\hat y^{\mathrm{collect}}(s),\overline M_{14}(s)\bigr),
\label{eq:choice}
\end{equation}
where lexicographic maximization first selects the highest predicted collect count and then resolves ties by mean cumulative simulated response. Other objectives order the quantities in $\mathcal E$ differently: an objective centered on population response ranks candidates by $\overline M_{14}$ or by response per CNY, $\overline M_{14}/b$.

\section{Evaluation Protocol}
\label{sec:evaluation}

Each evaluation examines one computation used in campaign comparison with data that can measure it: observed engagement of private RedNote notes tests the note predictors $g_\theta$; the controlled campaign traces how actions pass through the rollout $R_\phi$; public logs with randomized exposure, recorded assignment probabilities, or randomized treatment test policy-value estimators for content allocation and treatment-effect learners for audience ranking; and paired synthetic outcomes test feature-substitution scoring.

\subsection{Prediction and Controlled Response}
\label{subsec:prediction-evaluation}

We report the coefficient of determination $R^2_{\log}$ separately for reads, likes, collects, and comments after applying $\log(1+y)$ to observed and predicted counts. Shared folds compare predictor configurations; structured holdouts test temporal, creator, and niche generalization. Controlled-campaign rollouts record reach, mean content matching, engagement probability, and paid and organic response.

\subsection{Off-Policy Value Estimation}
\label{subsec:policy-evaluation}

A policy $\pi(a\mid x)$ assigns content or actions $a$ in context $x$; its value is the expected outcome under that assignment. Given logged outcome $y_i$ and logging probability $\mu_i$, define $w_i=\pi(a_i\mid x_i)/\mu_i$. Inverse propensity scoring (IPS) estimates the value by $\frac1n\sum_iw_iy_i$ and self-normalized IPS (SNIPS) by $\sum_iw_iy_i/\sum_iw_i$; the effective sample size $n_{\mathrm{eff}}=(\sum_iw_i)^2/\sum_iw_i^2$ describes weight concentration. Doubly robust (DR) estimation combines outcome predictions and weighted residuals; Switch-DR applies the residual correction below a weight threshold \citep{li2011unbiased,swaminathan2015self,dudik2011doubly,wang2017optimal}. Appendix~\ref{app:ope} specifies these estimators and their identification conditions.

\subsection{Audience Ranking and Paired Outcomes}
\label{subsec:ranking-evaluation}

For advertiser-randomized interventions, the conditional average treatment effect (CATE) is $\tau(x)=\mathbb E[Y(1)-Y(0)\mid X=x]$, where $Y(1)$ and $Y(0)$ are outcomes with and without intervention. We rank customers by estimated incremental response using treatment-effect learners \citep{kunzel2019metalearners,nie2021quasi,athey2019generalized}. For the top fraction $q$, uplift@$q$ is the treatment-minus-control difference in mean observed outcomes, $\bar y(T_q^{(1)})-\bar y(T_q^{(0)})$, where $T_q^{(1)}$ and $T_q^{(0)}$ are the treatment and control customers in the selected set. Normalized area under the uplift curve (AUUC) and Qini summarize performance across coverage fractions (Appendix~\ref{app:uplift}).

The paired synthetic test evaluates factual fit, alternative-outcome error, effect direction, and grouped CATE fit for fixed predictors under feature substitution (Section~\ref{sec:pairedexp}).

\section{Experiments}
\label{sec:experiments}

OranSim uses four LightGBM regressors for note engagement, a logistic response model for individual click and engagement probabilities, and a 60-segment Hawkes process for propagation. The prediction experiments fit the engagement regressors; the campaign case holds the fitted predictor and response mechanisms fixed while changing actions. Public-data experiments use off-policy estimators and treatment-effect learners for content allocation and audience selection. The paired synthetic experiment fits four more LightGBM regressors to score alternatives with known outcomes.

\subsection{RedNote Engagement Prediction}
\label{sec:predictionexp}

The five-fold comparison uses 39,000 private RedNote note snapshots and fits a LightGBM regressor for each of read, like, collect, and comment counts, with $\log(1+y)$ targets. The configurations without and with note age, v3 and v3.1-pg, share shuffled folds with seed 42. Titles and bodies are embedded separately with OpenAI text-embedding-3-small, each truncated to 768 dimensions; the concatenated 1,536 dimensions are reduced to 128 by PCA; adding metadata gives 182 features without note age and 183 with it.

The structured holdouts use a separately assembled sample of 12,154 notes from 13 niches, each with at least 100 reads. Each split refits 300-tree LightGBM regressors using 768-dimensional deterministic hashed representations of titles and bodies, concatenated and reduced to at most 64 dimensions by training-set PCA. The temporal split uses 10,331/1,823 training/test notes; the creator split uses 10,310/1,844, with 1,642 test creators absent from training; leave-one-niche-out retrains on 13 folds. Appendix~\ref{app:prediction} gives regressor settings, features, splits, and errors.

\begin{table}[t]
\centering
\small
\setlength{\tabcolsep}{3pt}
\caption{RedNote prediction measured by $R^2_{\log}$. The first two rows share notes and folds. Structured holdouts refit predictors on a separate corpus with hashed text features.}
\label{tab:prediction}
\begin{tabularx}{\linewidth}{@{}>{\raggedright\arraybackslash}X>{\raggedright\arraybackslash}p{2.05cm}rrrr@{}}
\toprule
Evaluation setting & Evaluation size & Reads & Likes & Collects & Comments \\
\midrule
Five-fold CV, without age (v3) & 39,000 & 0.5752 & 0.6122 & 0.5766 & 0.5492 \\
Five-fold CV, with age (v3.1-pg) & 39,000 & \textbf{0.5928} & \textbf{0.6223} & \textbf{0.5869} & \textbf{0.5585} \\
Latest-time holdout & 1,823 test notes & 0.253 & 0.361 & 0.329 & $-0.007$ \\
Unseen-creator holdout & 1,844 test notes & 0.559 & 0.550 & 0.513 & 0.347 \\
Leave-one-niche-out & 12,154 / 13 folds & 0.503 & 0.513 & 0.456 & 0.333 \\
\bottomrule
\end{tabularx}
\end{table}

Table~\ref{tab:prediction} shows that note age improves all four outcomes: the gains in unrounded $R^2_{\log}$ are 0.0176, 0.0101, 0.0103, and 0.0092 for reads, likes, collects, and comments. The resulting scores of 0.5585--0.6223 support engagement prediction for comparing candidate options. The unseen-creator scores of 0.347--0.559 and leave-one-niche-out scores of 0.333--0.513 support prediction for creators and niches absent from training. The latest-time scores are 0.253, 0.361, 0.329, and $-0.007$, respectively.

\subsection{Simulated Social Response to Marketing Actions}
\label{sec:case}

The controlled skincare campaign follows an action change through exposure, individual response, and propagation. We sample 100,000 individuals representing 2,000,000 people, fix the population with seed 2027, and compare five options with 30 shared seeds $\{0,\ldots,29\}$. Note predictions use the frozen v3.1-pg LightGBM predictor with its training text encoder at an evaluation age of 14 days. Response parameters stay fixed, and propagation uses $\beta=0.9$ per day, $r=0.35$, and six-hour steps (Appendix~\ref{app:simulator}).

The baseline combines informational creative A, a mid-tier creator, and CNY~40,000. Creative B presents the same triple-ceramide repair routine as a response to seasonal redness and invites viewers to take a skin-type test and claim a trial-size sample. Both are 30-second videos. Candidate options change the creative, budget, creator tier, or creative and budget together (Table~\ref{tab:case}). Targeting, platform, publication time, and exposure configuration are shared; Appendix~\ref{app:case} gives the full campaign.

\begin{table}[t]
\centering
\small
\setlength{\tabcolsep}{3pt}
\caption{Controlled campaign options. Inputs give creative, budget in CNY, and creator tier. Reads, likes, collects, and comments are predicted note counts at age 14 days; $\overline M_{14}$ is mean cumulative simulated response mass over 30 seeds.}
\label{tab:case}
\begin{tabular}{@{}llrrrrr@{}}
\toprule
Option & Inputs & Reads & Likes & Collects & Comments & $\overline M_{14}$ \\
\midrule
$s_0$ & A / 40,000 / mid & 54,819 & 2,209 & 1,371 & 52 & 1,834,405 \\
$s_c$ & B / 40,000 / mid & 56,520 & 2,016 & 1,020 & 80 & 1,832,978 \\
$s_b$ & A / 80,000 / mid & 54,819 & 2,209 & 1,371 & 52 & 3,599,734 \\
$s_k$ & A / 40,000 / high & 60,882 & 2,672 & 1,480 & 79 & 1,833,903 \\
$s_{cb}$ & B / 80,000 / mid & 56,520 & 2,016 & 1,020 & 80 & 3,604,845 \\
\bottomrule
\end{tabular}
\end{table}

\begin{table}[t]
\centering
\small
\setlength{\tabcolsep}{4pt}
\caption{The response process behind Table~\ref{tab:case}. Reach counts sampled individuals; match and engagement are means among those reached; $\overline M_{14}$ is scaled to the represented population. The last two columns give the percentage by which $\overline M_{14}$ exceeds two equal-reach controls, uniform selection and mean individual features. Means over the 30 shared seeds; Appendix~\ref{app:case-process} adds standard deviations.}
\label{tab:case-process}
\begin{tabular}{@{}lrrrrrr@{}}
\toprule
Option & Sample reach & Content match & Engagement prob. & $\overline M_{14}$ & vs.\ uniform & vs.\ mean \\
\midrule
$s_0$ & 41,666 & 0.5009 & 0.7020 & 1,834,405 & +6.61\% & $-0.17$\% \\
$s_c$ & 41,666 & 0.4971 & 0.7009 & 1,832,978 & +6.19\% & $-0.16$\% \\
$s_b$ & 83,333 & 0.4942 & 0.6783 & 3,599,734 & +4.59\% & $-0.19$\% \\
$s_k$ & 41,666 & 0.5011 & 0.7015 & 1,833,903 & +6.59\% & $-0.17$\% \\
$s_{cb}$ & 83,333 & 0.4987 & 0.6799 & 3,604,845 & +4.42\% & $-0.17$\% \\
\bottomrule
\end{tabular}
\end{table}

Increasing the budget expands reach and changes the composition of the reached audience. With creative A, doubling the budget selects 83,333 individuals in decreasing exposure-score order, up from 41,666. Mean content match falls from 0.5009 to 0.4942, and mean engagement probability from 0.7020 to 0.6783 (Table~\ref{tab:case-process}). Their initial responses enter the paid schedule and excite subsequent organic responses. Under the shared propagation parameters, $\overline M_{14}$ rises from 1,834,405 to 3,599,734: reach grows to 2.00 times the baseline and response mass to 1.96 times. The added individuals rank lower in exposure score and have lower mean response propensity, so response mass grows slightly less than reach.

Two equal-reach controls locate these differences (Table~\ref{tab:case-process}). Uniform selection of the same number of individuals lowers the mean engagement probability for $s_0$ from 0.7020 to 0.6346, and the full rollout exceeds it by 4.4\%--6.6\% in $\overline M_{14}$; giving every selected individual the audience's mean features changes $\overline M_{14}$ by less than 0.2\%. Heterogeneity thus shapes aggregate response mainly through which consumers the exposure ranking selects. With fixed initial segment responses, uniform cross-segment weights lower the targeted segments' organic share from 96.1\% under $A$ to 94.3\%, with total response unchanged (Appendix~\ref{app:case-mechanisms}).

At CNY~40,000, replacing A with B preserves the reach of 41,666 individuals and changes mean match from 0.5009 to 0.4971, engagement probability from 0.7020 to 0.7009, and response mass from 1,834,405 to 1,832,978, a paired difference of $-1{,}427$ (95\% bootstrap interval $[-1{,}510,-1{,}343]$). The historical-note predictor gives A 1,371 collects and B 1,020; B has more predicted reads and comments and fewer likes. The high-tier creator raises all four predictions, to 1,480 collects and 79 comments.

\begin{figure}[t]
\centering
\begin{subfigure}[t]{0.49\linewidth}
\centering
\includegraphics[width=\linewidth]{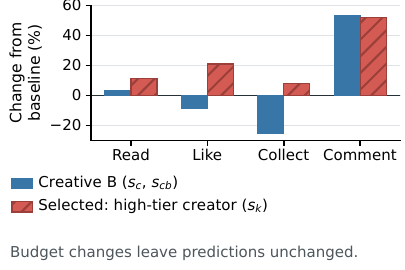}
\caption{Predicted changes from the baseline.}
\end{subfigure}\hfill
\begin{subfigure}[t]{0.49\linewidth}
\centering
\includegraphics[width=\linewidth]{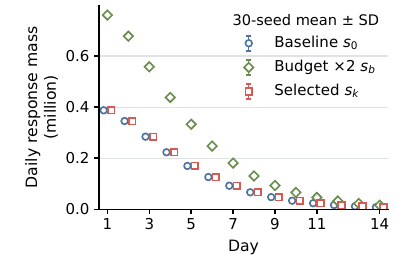}
\caption{Daily response of three options.}
\end{subfigure}
\caption{Campaign comparison in the synthetic skincare setting. (a) Changes in four note predictions; budget changes leave them unchanged, so $s_b$ matches the baseline and $s_{cb}$ matches creative B. (b) Daily simulated response, mean $\pm1$ standard deviation over 30 seeds; the nearly coinciding $s_0$ and $s_k$ points are offset horizontally. The collect-first rule selects the high-tier creator at CNY~40,000.}
\label{fig:case}
\end{figure}

Under the CNY~80,000 cap, the collect-first rule in Equation~\eqref{eq:choice} selects $s_k$, which has the most predicted collects. Options that differ only in budget share their predictions, and cumulative response mass orders each such pair, placing $s_b$ ahead of $s_0$ and $s_{cb}$ ahead of $s_c$. The comparison thus connects the engagement objective to creator and creative choice and the population response to budget choice. Under objectives centered on population response, the rollout decides: ranking by $\overline M_{14}$ selects $s_{cb}$ and ranking by response per CNY selects $s_0$, as doubling the budget raises response mass by 96.2\% and lowers response per CNY by 1.9\%. All three rankings hold under eleven single-parameter changes to propagation, targeting, exploration, and response parameters, which keep the budget-doubling ratio within 1.950--1.973 (Appendix~\ref{app:case-sensitivity}). Figure~\ref{fig:case} shows the predictions and daily responses. Repeated same-seed baselines have zero aggregate and daily differences (Appendix~\ref{app:case-repro}).

\subsection{Policy Value and Audience Ranking on Public Data}
\label{sec:publicexp}

We evaluate content allocation as policy-value estimation and audience selection as treatment-effect ranking. KuaiRand-Pure supplies 1,186,059 randomized exposure records from 27,285 users and 7,583 videos, with logging probability $1/7{,}583$ and long view as the outcome \citep{gao2022kuairand}. Open Bandit supplies 10,000 records each from random and Bernoulli Thompson Sampling (BTS) policies over 80 actions \citep{saito2020open}. Each log is split in time: the earlier part selects the ten most-logged actions, shared by the two Open Bandit logs, and fits the reward means used by DR and Switch-DR; the later part is evaluated. The target policy allocates probability 0.1 uniformly over all actions and 0.9 uniformly over the ten. Intervals come from 1,000 bootstrap resamples of users on KuaiRand and of timestamps on Open Bandit (Appendix~\ref{app:ope}).

X5 RetailHero contains 200,039 customers, treatment fraction 0.4998, and outcome rate 0.6199. Seeds $\{42,137,256\}$ repeat training with shared 75\%/25\% treatment/outcome-stratified splits: 150,029 training and 50,010 test customers per seed. The methods are random ranking, two-model logistic regression, S/T/X learners and DRLearner with 80-tree forests, and CausalForestDML with 200 causal trees; Appendix~\ref{app:x5} gives configurations and every reach fraction.

On KuaiRand, SNIPS is 0.0879 with interval $[0.0755,0.1035]$, and IPS, DR, and Switch-DR lie within 0.0828--0.0932. The target policy concentrates probability on ten videos, so their records carry weight 682.57 and all others 0.10; these heavily weighted records reduce the effective sample size to 1,454 of 847,973 records. The mean weight is 1.060 with interval $[1.005,1.114]$. On the random and BTS Open Bandit logs, with 19 and 17 evaluated clicks, SNIPS is 0.0100 with interval $[0.0034,0.0184]$ and 0.0022 with $[0.0007,0.0040]$ (Table~\ref{tab:public}).

On X5, two-model logistic regression has the highest mean AUUC and X-Learner the highest Qini; DRLearner leads in uplift at 5\% and 10\% reach and X-Learner at 20\% and 30\%, so the preferred ranking method depends on the reachable fraction of customers (Appendix~\ref{app:x5}).

\subsection{Paired Counterfactual Scoring}
\label{sec:pairedexp}

We test feature-substitution scoring using 20,000 synthetic scenarios with known factual and alternative outcomes, split by scenario identifier into 16,000/2,000/2,000 training/validation/test cases. The alternative setting changes the budget by a prespecified rule while preserving creator attributes. The generator computes exposure, clicks, conversions, and revenue under each setting (Appendix~\ref{app:paired-data}).

Four 300-tree LightGBM regressors fit the factual training outcomes. Each test scenario is scored twice with fixed predictor parameters, substituting the factual and alternative features; their difference estimates the intervention effect (Table~\ref{tab:paired}). CATE evaluation averages generated and predicted differences within budget tiers, creator tiers, and niches before computing $R^2$; within-group averaging reduces the influence of the independently drawn outcome perturbations. Appendix~\ref{app:paired} gives all metrics.

\begin{table}[t]
\centering
\small
\setlength{\tabcolsep}{5pt}
\caption{LightGBM feature-substitution scoring on 2,000 paired synthetic test scenarios. Counterfactual normalized MAE (nMAE) divides MAE by the mean absolute alternative outcome.}
\label{tab:paired}
\begin{tabular}{@{}lrrrr@{}}
\toprule
Outcome & Factual $R^2$ & Counterfactual nMAE & Effect-sign accuracy & Grouped CATE $R^2$ \\
\midrule
Exposure & 0.964 & 0.219 & 0.955 & 0.457 \\
Clicks & 0.892 & 0.256 & 0.617 & 0.355 \\
Conversions & 0.762 & 0.582 & 0.930 & 0.531 \\
Revenue & 0.667 & 0.439 & 0.552 & 0.003 \\
\bottomrule
\end{tabular}
\end{table}

\section{Limitations}
\label{sec:limitations}

The outcome predictors fit a private observational RedNote corpus selected by a minimum engagement threshold and within-niche engagement rankings. They model the platform distribution represented by this corpus, and their prediction differences between creatives or creators reflect associations observed in these notes. Back-transformed counts fall 57\%--82\% below observed means in the structured holdouts, and the collect-first rule uses only their order (Appendix~\ref{app:prediction-counts}). Training note age is counted to the fitting date rather than to each count's collection, which changes temporal errors (Appendix~\ref{app:prediction-age}). Retraining the predictors requires the private corpus and the text-embedding-3-small service; the deterministic snapshot checks scoring with hashed text features.

The four note-outcome predictors are fitted to observations; response, propagation, and intergroup influence use design parameters, which also set absolute response levels, so the case compares options by differences. Calibrating these parameters and testing intervention differences and trajectory shapes on real platforms require campaign-level observations. Rollout content matching uses hashed titles without semantic relations to interests. X5 supplies randomized treatment labels and observed responses for ranking evaluation; individual conditional treatment effects remain unobservable.

\section{Conclusion}
\label{sec:conclusion}

OranSim connects marketing actions to population response through exposure allocation, heterogeneous individual behavior, and intergroup propagation. Shared initial populations and aligned random numbers allow candidate campaigns to be compared along this process. The controlled campaign shows how expanding reach changes audience composition and cumulative response; historical-note predictors supply engagement estimates for objective-specific selection. The prediction, public-data, and paired synthetic experiments evaluate the predictors and estimators used in these comparisons. Together, this representation and evaluation support prelaunch analysis of how a campaign reaches consumers and how their responses accumulate over time.

\paragraph{AI-assisted manuscript preparation.}
The research ideas, paper outline, and key arguments were developed by the human authors. OpenAI Codex assisted with refining the manuscript's organization and wording and with the English translation.

\paragraph{Reproducibility statement.}
Appendix~\ref{app:simulator} specifies the population, exposure and response equations, propagation parameters, and rollout procedure. Appendix~\ref{app:case} records the campaign settings, shared-seed results, mechanism controls, and parameter sensitivity. Appendices~\ref{app:prediction}--\ref{app:paired} describe the prediction splits, count-scale errors, note-age reference, policy estimators, ranking metrics, and synthetic experiment. Appendix~\ref{app:snapshot} documents the deterministic scoring snapshot, and Appendix~\ref{app:materials} lists the released code and results; the public-data and paired synthetic experiments can be rerun from public data, the released generator, and the released scripts.

\FloatBarrier
\setlength{\bibsep}{1pt plus 0.5pt}
\bibliographystyle{plainnat}
\bibliography{references}

\clearpage
\appendix
\counterwithin{table}{section}
\counterwithin{figure}{section}
\counterwithin{equation}{section}

\FloatBarrier
\section{Simulator Specification}
\label{app:simulator}

This appendix gives the parameters and steps of the computation described in Sections~\ref{subsec:society}--\ref{subsec:scenarios}. Table~\ref{tab:app-parameters} collects all constants for reference.

\subsection{Simulated Population and Segments}
\label{app:population}

The simulated population contains 100,000 sampled individuals representing 2,000,000 people, so the population scaling weight is $w_{\mathcal P}=20$. The population is generated once with seed 2027 and shared by all candidate scenarios. Each individual has an age band, gender, city tier, income decile, education level, and occupation; a 64-dimensional unit interest vector; Big Five trait scores in $[0,1]$; an activity level on each platform; and a 16-dimensional response state. Age band, gender, and city tier are drawn from marginal proportions, occupation is drawn conditional on age band, and income is correlated with city tier and education.

The combinations of six age bands (15--24, 25--34, 35--44, 45--54, 55--64, and 65 or older), two genders, and five city tiers (tier 1 through tier 5 and below) define the $6\times2\times5=60$ population segments. Exposure and individual response are computed for individuals; propagation is computed between segments.

\subsection{Exposure Scoring and Budget Conversion}
\label{app:exposure}

On platform $p$, the exposure score of individual $i$ is the product of six factors,
\begin{equation}
e_i=m_i^{\mathrm{content}}\,a_{ip}\,g_i^{\mathrm{target}}\,g_i^{\mathrm{creator}}\,\pi_{ip}\,\xi_i .
\label{eq:app-exposure}
\end{equation}
Content matching $m_i^{\mathrm{content}}=(\langle\mathbf u_i,\mathbf c\rangle+1)/2$ maps the inner product of the individual's interest vector $\mathbf u_i$ and the creative's content vector $\mathbf c$ to $[0,1]$. The content vector is a deterministic hashed representation of the creative title, biased for a small set of gender-, age-, and city-related keywords and then normalized; neither title in the case study contains these keywords. The activity level $a_{ip}$ is the individual's use frequency on platform $p$. The targeting weight $g_i^{\mathrm{target}}$ equals the targeting strength 2.0 when the individual satisfies any condition listed in the audience setting (age band, gender, or city tier) and $1/2.0$ otherwise. Creator matching $g_i^{\mathrm{creator}}=1+0.5\,\mathrm{clip}(\langle\mathbf u_i,\mathbf k\rangle,0,1)$ measures how close the individual's interests are to the creator's audience vector $\mathbf k$; when the creator's niche has a follower-profile prior, it is further multiplied by a gender-, age-, city-, and income-based weight normalized to mean one. The creator niche in the case study has no such prior, so this weight equals one. On RedNote, the platform audience prior $\pi_{ip}$ multiplies 1.33 for women and 0.68 for men, 1.8 for tier-1 and tier-2 cities and 0.6 for tier-3 cities and below, and 1.4 for ages 15--44 and 0.6 for ages 55 and older. The exploration perturbation $\xi_i\sim\mathcal U(1-0.4\delta_p,\,1+0.4\delta_p)$ is drawn with the scenario seed, with exploration strength $\delta_p=0.65$ on RedNote.

The budget determines how many individuals are selected. The platform budget on the simulated sample is $b\rho_p/w_{\mathcal P}$, and the nominal number of impressions is $1000\,b\rho_p/(w_{\mathcal P}\,\mathrm{CPM}_p)$, where the cost per thousand impressions on RedNote is CNY~48. OranSim selects individuals in decreasing order of $e_i$, up to the nominal number of impressions or the full sample size, whichever is smaller. For the CNY~40,000 budget of the case study, the sample budget is CNY~2,000 and the nominal number of impressions is 41,666.7, so 41,666 individuals are selected; scaled by $w_{\mathcal P}$, this corresponds to 833,333 nominal impressions and a represented reach of 833,320 people.

\subsection{Individual Response and Paid Schedule}
\label{app:response}

For each selected individual, the click and engagement probabilities are
\begin{equation}
\begin{aligned}
p_i^{\mathrm{click}}&=\sigma\!\left(\mathbf w_{\mathrm c}^\top\mathbf x_i^{\mathrm c}-1.2-\kappa_i+0.7\,\epsilon_i\right),\\
p_i^{\mathrm{engage}}&=p_i^{\mathrm{click}}\,\sigma\!\left(\mathbf w_{\mathrm e}^\top\mathbf x_i^{\mathrm e}-1.5+0.5\,\epsilon_i\right).
\end{aligned}
\label{eq:app-response}
\end{equation}
The click features $\mathbf x_i^{\mathrm c}$ are content matching, platform activity, targeting weight, creator matching, a fatigue proxy, openness, and a celebrity indicator, with weights $\mathbf w_{\mathrm c}=(1.8,1.2,0.9,0.4,-0.7,0.25,0.3)$; the fatigue proxy is the absolute value of the first response-state coordinate, clipped to $[0,1]$. The engagement features $\mathbf x_i^{\mathrm e}$ are content matching, platform activity, targeting weight, creator matching, neuroticism, and openness, with weights $\mathbf w_{\mathrm e}=(1.4,0.9,0.6,0.5,-0.3,0.4)$. The term $\kappa_i$ collects click penalties from compliance risk and AIGC labeling of the creative; both are zero for the two creatives in the case study. The latent response perturbation $\epsilon_i\sim\mathcal N(0,1)$ is drawn for every sampled individual with seed $\omega+71{,}000$, so an individual uses the same perturbation in every scenario under a given seed. The mean engagement probability reported in the main text is the mean of $p_i^{\mathrm{engage}}$ over selected individuals.

The initial response mass entering propagation is $p_i^{\mathrm{click}}+0.5\,p_i^{\mathrm{engage}}$. Individual masses are summed within segments and then distributed over days by the paid schedule: the share of day $d$ is proportional to $\exp[-0.4(d-1)]$, normalized over 14 days, and divided equally among four six-hour steps. Under this schedule, day~1 receives 33.1\% of the paid response mass and the first two days receive 55.3\%.

\subsection{Segment-Level Propagation}
\label{app:propagation}

In Equation~\eqref{eq:dynamics}, the decay rate is $\beta=0.9$ per day, the branching parameter is $r=0.35$, and the step size is $\Delta t=0.25$ days. The intergroup influence matrix $A\in\mathbb R^{60\times60}$ is built from segment attributes. The unnormalized weight is 3.0 within a segment and, between segments $i\neq j$,
\begin{equation}
\tilde A_{ij}=0.15\exp(-0.5\,d_{ij}),\qquad
d_{ij}=|\Delta_{\mathrm{age}}|+0.8\,\mathbb 1[\text{genders differ}]+0.6\,|\Delta_{\mathrm{city}}|,
\label{eq:app-influence}
\end{equation}
where $\Delta_{\mathrm{age}}$ and $\Delta_{\mathrm{city}}$ are differences in age-band and city-tier indices. Each row is then divided by its sum, so the influence weights of every source segment sum to one. The construction makes interpersonal influence stronger between similar people: influence is largest within a segment and weakens as age, gender, and city tier differ more. Studies of interpersonal communication find that people receive information and opinions more readily from others like themselves \citep{katz1955personal}, and agent-based marketing models represent such influence through local connections between consumers \citep{rand2011agent}.

Because every row of $A$ sums to one, summing Equation~\eqref{eq:dynamics} over segments gives a recursion for the total excitation $z_\tau=\mathbf 1^\top\mathbf S_\tau$:
\begin{equation}
z_{\tau+1}=\bigl(e^{-\beta\Delta t}+r\Delta t\bigr)z_\tau+\mathbf 1^\top\boldsymbol\mu_\tau .
\label{eq:app-total}
\end{equation}
With the values above, $e^{-\beta\Delta t}+r\Delta t=0.886<1$, so total excitation decays geometrically after the paid input ends. The total organic response is therefore determined by the paid input and the parameters $r$, $\beta$, and $\Delta t$, and $A$ determines how it is distributed across segments.

\subsection{Rollout Procedure}
\label{app:rollout}

For a scenario $s$ and seed $\omega$, the rollout operator $R_\phi$ performs the following steps.
\begin{enumerate}
  \item \textbf{Shared population.} Use the population of Appendix~\ref{app:population}; the scaling weight $w_{\mathcal P}$ and the scenario's budget and platform shares give the sample budget.
  \item \textbf{Exposure.} Draw the exploration perturbation with seed $\omega$, compute exposure scores by Equation~\eqref{eq:app-exposure}, and select individuals.
  \item \textbf{Individual response.} Read the latent perturbations $\epsilon_i$ under seed $\omega+71{,}000$ and compute click probabilities, engagement probabilities, and initial response masses by Equation~\eqref{eq:app-response}.
  \item \textbf{Paid injection.} Sum initial response masses within segments and distribute them over 56 steps by the paid schedule, giving $\boldsymbol\mu_\tau$.
  \item \textbf{Propagation.} Starting from $\mathbf S_0=\mathbf 0$, compute the organic response rate $\boldsymbol\lambda_\tau$ and the state $\mathbf S_{\tau+1}$ step by step by Equation~\eqref{eq:dynamics}.
  \item \textbf{Aggregation.} Merge the four steps of each day into daily paid and organic response masses, scale them by $w_{\mathcal P}$, and accumulate them into $M_{14}(s,\omega)$ by Equation~\eqref{eq:mass}; record the selected individuals, mean content matching, mean engagement probability, and daily trajectories.
\end{enumerate}
Scenarios differ only in the actions used in steps 2--6; the population, mechanism parameters, and per-seed latent perturbations stay fixed.

\begin{table}[t]
\centering
\small
\setlength{\tabcolsep}{4pt}
\caption{Simulator parameters.}
\label{tab:app-parameters}
\begin{tabularx}{\linewidth}{@{}>{\raggedright\arraybackslash}p{2.2cm}>{\raggedright\arraybackslash}p{4.6cm}>{\raggedright\arraybackslash}Xl@{}}
\toprule
Symbol & Value & Role & Section \\
\midrule
$N$, $w_{\mathcal P}$ & 100,000; 20 & Sample size and population scaling weight & \ref{app:population} \\
-- & 2027 & Population seed & \ref{app:population} \\
-- & $6\times2\times5=60$ & Population segments & \ref{app:population} \\
-- & 2.0; $1/2.0$ & Targeting strength (matched; unmatched) & \ref{app:exposure} \\
-- & $1+0.5\,\mathrm{clip}(\cdot)$ & Creator matching & \ref{app:exposure} \\
$\pi_{ip}$ & 1.33/0.68; 1.8/0.6; 1.4/0.6 & RedNote audience prior (gender; city tier; age) & \ref{app:exposure} \\
$\delta_p$ & 0.65 & RedNote exploration strength & \ref{app:exposure} \\
$\mathrm{CPM}_p$ & CNY 48 & RedNote cost per thousand impressions & \ref{app:exposure} \\
$\mathbf w_{\mathrm c}$; intercept & $(1.8,1.2,0.9,0.4,-0.7,0.25,0.3)$;\newline $-1.2$ & Click model & \ref{app:response} \\
$\mathbf w_{\mathrm e}$; intercept & $(1.4,0.9,0.6,0.5,-0.3,0.4)$;\newline $-1.5$ & Engagement model & \ref{app:response} \\
-- & 0.7; 0.5 & Latent perturbation scale (click; engagement) & \ref{app:response} \\
-- & 0.5 & Engagement weight in initial response mass & \ref{app:response} \\
-- & 0.4 & Daily decay of the paid schedule & \ref{app:response} \\
$\beta$ & 0.9 day$^{-1}$ & Excitation decay rate & \ref{app:propagation} \\
$r$ & 0.35 & Branching parameter & \ref{app:propagation} \\
$\Delta t$ & 0.25 day & Step size & \ref{app:propagation} \\
-- & 3.0; 0.15; (1, 0.8, 0.6) & Influence matrix: within-segment weight; cross-segment scale; age, gender, city distance coefficients & \ref{app:propagation} \\
\bottomrule
\end{tabularx}
\end{table}

\FloatBarrier
\section{Case Study: Controlled Skincare Campaign}
\label{app:case}

This appendix records the controlled case of Section~\ref{sec:case} in full. It follows the order of the main text: the frozen campaign specification, the reach, response tendencies, and cumulative response of all five options, their daily trajectories, the application of the collect-first rule, seed variation with reproducibility checks, mechanism controls, and parameter sensitivity.

\subsection{Campaign Specification}
\label{app:case-spec}

Table~\ref{tab:app-case-spec} lists the settings shared by all options. The audience setting lists conditions on age band, gender, and city tier, and targeting weights follow Appendix~\ref{app:exposure}. Both creatives use synthetic Chinese-language copy written by the authors; the table gives English translations. Table~\ref{tab:app-case-options} lists the five options: each single-factor option changes one field of the baseline, and the combined option changes the creative and the budget.

\begin{table}[t]
\centering
\small
\caption{Campaign settings shared by all options.}
\label{tab:app-case-spec}
\begin{tabularx}{\linewidth}{@{}>{\raggedright\arraybackslash}p{3.6cm}>{\raggedright\arraybackslash}X@{}}
\toprule
Field & Setting \\
\midrule
Platform and niche & RedNote; skincare \\
Publication time & 2026-03-16 10:00 \\
Outcome evaluation age $\gamma$ & 14 days \\
Audience setting & Age bands 15--24, 25--34, and 35--44; women; tier-1 to tier-3 cities; targeting strength 2.0 \\
Platform shares $\boldsymbol\rho$ & RedNote 100\% \\
Schedule and propagation & Appendices~\ref{app:response}--\ref{app:propagation} \\
Population and seeds & Shared population of Appendix~\ref{app:population}; seeds $\{0,\ldots,29\}$ \\
Creative A & Title: ``Spring repair for sensitive skin: a triple-ceramide formula and a 28-day usage guide.'' Body: introduces the formula, suitable skin types, and morning and evening steps, and points to the full ingredient list. Topics: sensitive skin, ceramides, spring skincare. 30-second video; minimal visuals; calm music. \\
Creative B & Title: ``Redness when the season changes? A 30-second look at a triple-ceramide repair routine.'' Body: invites viewers to take a skin-type test and claim a trial-size sample, and points to a staged 28-day routine. Topics: seasonal redness, sensitive skin, trial size. 30-second video; bright visuals; calm music. \\
Mid-tier creator & 300,000 followers; interaction rate 4.5\% \\
High-tier creator & 1,200,000 followers; interaction rate 3.8\% \\
Campaign objective & Budget cap CNY 80,000; compare predicted collects first and $\overline M_{14}$ for ties \\
\bottomrule
\end{tabularx}
\end{table}

\begin{table}[t]
\centering
\small
\caption{Campaign options and the fields they change relative to the baseline.}
\label{tab:app-case-options}
\begin{tabular}{@{}llrll@{}}
\toprule
Option & Creative & Budget (CNY) & Creator & Changed fields \\
\midrule
$s_0$, baseline & A & 40,000 & mid-tier & -- \\
$s_c$, creative change & B & 40,000 & mid-tier & creative \\
$s_b$, budget increase & A & 80,000 & mid-tier & budget \\
$s_k$, creator change & A & 40,000 & high-tier & creator \\
$s_{cb}$, creative and budget & B & 80,000 & mid-tier & creative, budget \\
\bottomrule
\end{tabular}
\end{table}

\subsection{Reach, Response Tendencies, and Cumulative Response}
\label{app:case-process}

Table~\ref{tab:app-case-process} lists the process quantities of the five options in the order of the rollout. Nominal impressions determine how many individuals are selected; the content matching and engagement probability of the selected individuals determine the initial response mass; the initial responses are injected according to the paid schedule, excite organic responses, and accumulate into the 14-day cumulative simulated response mass.

\begin{table*}[t]
\centering
\small
\setlength{\tabcolsep}{3.5pt}
\caption{Rollout quantities of the five options. Means $\pm$ standard deviations are over 30 shared seeds; reach is identical across seeds. Response masses are scaled by $w_{\mathcal P}$. Match and engagement are means over selected individuals.}
\label{tab:app-case-process}
\textit{(a) Exposure and individual response}\par\smallskip
\begin{tabular*}{\linewidth}{@{\extracolsep{\fill}}lccccc@{}}
\toprule
Option & \shortstack{Nominal\\impressions} & \shortstack{Sample\\reach} & \shortstack{Represented\\reach} & \shortstack{Content\\match} & \shortstack{Engagement\\probability} \\
\midrule
$s_0$ & 833,333 & 41,666 & 833,320 & $0.5009\pm0.0001$ & $0.7020\pm0.0007$ \\
$s_c$ & 833,333 & 41,666 & 833,320 & $0.4971\pm0.0000$ & $0.7009\pm0.0007$ \\
$s_b$ & 1,666,667 & 83,333 & 1,666,660 & $0.4942\pm0.0000$ & $0.6783\pm0.0004$ \\
$s_k$ & 833,333 & 41,666 & 833,320 & $0.5011\pm0.0001$ & $0.7015\pm0.0007$ \\
$s_{cb}$ & 1,666,667 & 83,333 & 1,666,660 & $0.4987\pm0.0000$ & $0.6799\pm0.0004$ \\
\bottomrule
\end{tabular*}
\par\medskip
\textit{(b) Cumulative response over 14 days}\par\smallskip
\begin{tabular*}{\linewidth}{@{\extracolsep{\fill}}lccc@{}}
\toprule
Option & \shortstack{14-day paid\\response} & \shortstack{14-day organic\\response} & $\overline M_{14}$ \\
\midrule
$s_0$ & $1{,}043{,}738\pm565$ & $790{,}668\pm428$ & $1{,}834{,}405\pm992$ \\
$s_c$ & $1{,}042{,}926\pm589$ & $790{,}053\pm446$ & $1{,}832{,}978\pm1{,}034$ \\
$s_b$ & $2{,}048{,}172\pm738$ & $1{,}551{,}562\pm559$ & $3{,}599{,}734\pm1{,}296$ \\
$s_k$ & $1{,}043{,}452\pm585$ & $790{,}451\pm443$ & $1{,}833{,}903\pm1{,}029$ \\
$s_{cb}$ & $2{,}051{,}080\pm739$ & $1{,}553{,}765\pm560$ & $3{,}604{,}845\pm1{,}299$ \\
\bottomrule
\end{tabular*}
\end{table*}

Options with the same budget have the same nominal impressions and reach, and their process quantities differ through the composition of the selected individuals. At CNY~40,000, the creator change moves the mean content match from 0.5009 to 0.5011 and the mean engagement probability from 0.7020 to 0.7015, with a cumulative simulated response mass of 1,833,903. Doubling the budget increases the number of selected individuals from 41,666 to 83,333. Under creative B, raising the budget from CNY~40,000 to CNY~80,000 moves the mean content match from 0.4971 to 0.4987, lowers the mean engagement probability from 0.7009 to 0.6799, and raises the cumulative simulated response mass from 1,832,978 to 3,604,845.

\subsection{Daily Paid and Organic Response}
\label{app:case-daily}

\begin{figure}[t]
\centering
\includegraphics[width=\linewidth]{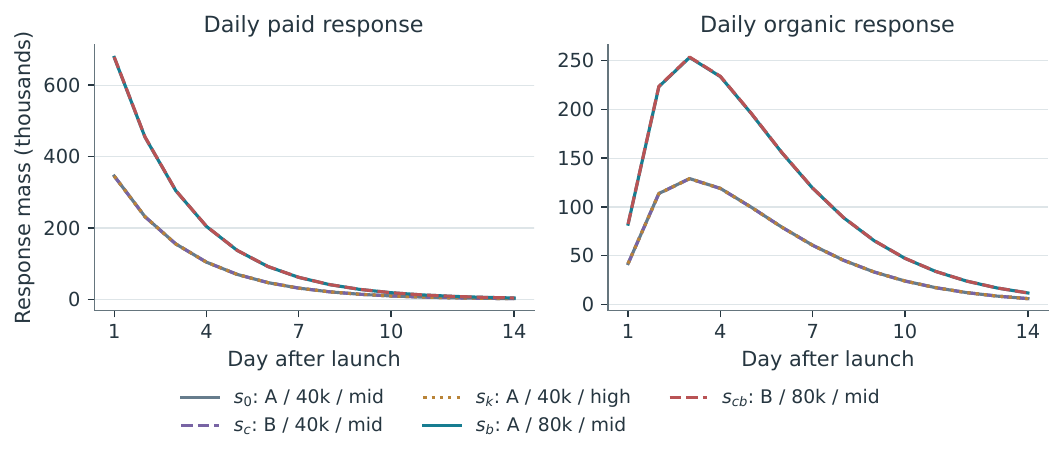}
\caption{Daily paid (left) and organic (right) response of the five options. Lines are means over 30 shared seeds and bands show $\pm1$ standard deviation.}
\label{fig:app-case-daily}
\end{figure}

Figure~\ref{fig:app-case-daily} shows the daily paid and organic responses of the five options. Paid response is highest on day~1 under the paid schedule and declines thereafter; organic response accumulates from the excitation of earlier responses, peaks on day~3, and then decays. Options with the same budget nearly coincide at the plotted scale, and on every day the paid and organic responses of the two CNY~80,000 options are about 1.96 times those of the CNY~40,000 options. For $s_0$, the paid and organic responses are 345,376 and 41,984 on day~1, 155,188 and 129,033 on day~3, and 1,905 and 6,041 on day~14.

\subsection{Applying the Collect-First Rule}
\label{app:case-rule}

Equation~\eqref{eq:choice} is applied in three steps. The budget cap of CNY~80,000 keeps all five options. Ranking by predicted collects gives 1,480 for $s_k$, 1,371 for $s_0$ and $s_b$, and 1,020 for $s_c$ and $s_{cb}$; the budget is not a predictor input, so options with the same creative and creator but different budgets have identical predictions. Within each tied pair, the rule compares $\overline M_{14}$: $s_b$ precedes $s_0$ (3,599,734 versus 1,834,405), and $s_{cb}$ precedes $s_c$ (3,604,845 versus 1,832,978). The complete order is $s_k$, $s_b$, $s_0$, $s_{cb}$, $s_c$, and the rule selects $s_k$. Table~\ref{tab:app-case-deltas} lists the change of each prediction relative to the baseline: the creator change raises all four predictions, and the creative change raises reads and comments while lowering likes and collects.

\begin{table}[t]
\centering
\small
\setlength{\tabcolsep}{4pt}
\caption{Change in predicted engagement counts relative to the baseline $s_0$, with relative changes in parentheses.}
\label{tab:app-case-deltas}
\begin{tabular}{@{}lrrrr@{}}
\toprule
Option & Reads & Likes & Collects & Comments \\
\midrule
$s_c$, $s_{cb}$ & +1,701 (+3.1\%) & $-193$ ($-8.7\%$) & $-350$ ($-25.6\%$) & +28 (+53.2\%) \\
$s_k$ & +6,063 (+11.1\%) & +463 (+21.0\%) & +109 (+8.0\%) & +27 (+52.0\%) \\
$s_b$ & 0 & 0 & 0 & 0 \\
\bottomrule
\end{tabular}
\end{table}

\subsection{Seed Variation and Reproducibility Checks}
\label{app:case-repro}

Across the 30 shared seeds, the standard deviation of $M_{14}$ lies between 992 and 1,299 for the five options. Differences between options are computed within each seed, where the population and latent response noise are shared. Table~\ref{tab:app-case-paired} reports the mean paired differences with percentile intervals from 10,000 bootstrap resamples of the 30 seeds, drawn jointly for all options. The creative effect changes sign with the budget: creative B lowers $M_{14}$ at CNY~40,000 and raises it at CNY~80,000. Relative differences divide by the comparator's mean, and per-CNY differences first divide each option's $M_{14}$ by its budget. The intervals describe simulation randomness for the fixed population, mechanism parameters, and note predictions.

\begin{table}[t]
\centering
\small
\setlength{\tabcolsep}{4pt}
\caption{Paired differences in $M_{14}$ over the 30 shared seeds, with 95\% bootstrap intervals.}
\label{tab:app-case-paired}
\begin{tabular}{@{}lrrrr@{}}
\toprule
Contrast & Mean difference & 95\% interval & Relative & Per CNY \\
\midrule
$s_c-s_0$ & $-1{,}427$ & $[-1{,}510,\,-1{,}343]$ & $-0.08$\% & $-0.08$\% \\
$s_k-s_0$ & $-502$ & $[-548,\,-456]$ & $-0.03$\% & $-0.03$\% \\
$s_b-s_0$ & $1{,}765{,}329$ & $[1{,}764{,}964,\,1{,}765{,}680]$ & $+96.23$\% & $-1.88$\% \\
$s_{cb}-s_b$ & $5{,}111$ & $[5{,}056,\,5{,}168]$ & $+0.14$\% & $+0.14$\% \\
$(s_{cb}-s_b)-(s_c-s_0)$ & $6{,}538$ & $[6{,}436,\,6{,}639]$ & $+0.36$\% & $+0.22$\% \\
\bottomrule
\end{tabular}
\end{table}

The frozen results include the following checks. Repeating the baseline with the same seed gives a maximum absolute difference of zero in both aggregate and daily outputs; an empty intervention that changes no field gives a zero difference; all five options share the same population hash; and all rollout summaries are finite. The scenario manifest records the fields each option changes relative to the baseline, matching Table~\ref{tab:app-case-options}.

The predictions use the v3.1-pg predictor with title and body representations from its training encoder, text-embedding-3-small (first 768 dimensions, renormalized). The four vectors are stored with the results, so the predictions are reproduced without API access. The principal component transformation in the predictor was serialized with scikit-learn 1.8.0 and emits a version warning when loaded in the 1.6.1 runtime; inference completes, and the determinism and finite-output checks above pass. The frozen results also record hashes of the campaign specification, model checkpoints, population, random-number configuration, and stored text representations.

\subsection{Mechanism Controls}
\label{app:case-mechanisms}

Two controls keep the number of selected individuals and each individual's latent response noise. Uniform selection replaces the exposure ranking by a fixed random order over individual identifiers and selects its first $K_p$ individuals, so the sets selected under different budgets are nested; each selected individual keeps its own features. The mean-feature control keeps the individuals selected by the full rollout and replaces each response feature, after its transformations, by its mean over the selected audience before computing click and engagement probabilities. Both controls use the same paid schedule and propagation. Besides the five options, three options extend the comparison: $s_l$ halves the baseline budget to CNY~20,000, and $s_t$ and $s_{tb}$ set every targeting weight to one at CNY~40,000 and CNY~80,000. The populations generated with seeds 2028 and 2029 repeat all comparisons.

Table~\ref{tab:app-mechanisms} reports the population of the main text. The 95\% paired bootstrap intervals of the differences, from 10,000 resamples of the 30 seeds, lie within $\pm0.04$ percentage points of each estimate, and each percentage changes by less than 0.07 percentage points across the three populations. The full rollout exceeds uniform selection by 1.1\%--8.5\%, with the largest margin at the smallest budget, where the ranking is truncated earliest. Its difference from the mean-feature control stays between $-0.21$\% and $-0.14$\%. With neutral targeting, $\overline M_{14}$ at CNY~40,000 falls from 1,834,405 to 1,561,241, although all selected individuals still meet at least one audience condition (97.6\% at CNY~80,000). The targeting weight enters both the exposure score and the click and engagement features (Appendix~\ref{app:response}), and setting it to one lowers the mean engagement probability from 0.7020 to 0.5368.

\begin{table}[t]
\centering
\small
\setlength{\tabcolsep}{4pt}
\caption{Full rollout versus the two equal-reach controls in the main-text population. Percentages give how much the full rollout's $\overline M_{14}$ exceeds each control, averaged over 30 seeds.}
\label{tab:app-mechanisms}
\begin{tabular}{@{}lrrrr@{}}
\toprule
Option & Budget (CNY) & $\overline M_{14}$ & vs.\ uniform selection & vs.\ mean features \\
\midrule
$s_0$ & 40,000 & 1,834,405 & $+6.61$\% & $-0.17$\% \\
$s_c$ & 40,000 & 1,832,978 & $+6.19$\% & $-0.16$\% \\
$s_b$ & 80,000 & 3,599,734 & $+4.59$\% & $-0.19$\% \\
$s_k$ & 40,000 & 1,833,903 & $+6.59$\% & $-0.17$\% \\
$s_{cb}$ & 80,000 & 3,604,845 & $+4.42$\% & $-0.17$\% \\
$s_l$ & 20,000 & 933,383 & $+8.49$\% & $-0.14$\% \\
$s_t$, neutral targeting & 40,000 & 1,561,241 & $+4.45$\% & $-0.20$\% \\
$s_{tb}$, neutral targeting & 80,000 & 3,021,485 & $+1.08$\% & $-0.21$\% \\
\bottomrule
\end{tabular}
\end{table}

The propagation comparison fixes, for every seed, the 60 segment injections $\boldsymbol\mu_\tau$ of an option and replaces only $A$. The within-segment matrix is the identity. The uniform cross-segment matrix keeps each diagonal weight of $A$ and spreads the remaining weight of its row equally over the other 59 segments. Table~\ref{tab:app-propagation} compares the daily distribution of organic response over segments with the distribution under $A$ by the total variation distance $\frac12\sum_g|p_g(d)-q_g(d)|$, averaged over the 14 days and 30 seeds. Total response differs from that under $A$ by at most $2\times10^{-7}$ in relative terms, the float32 rounding level, as Equation~\eqref{eq:app-total} implies. With within-segment propagation every segment has the same organic timing; under $A$, the time centroids of segment organic response range from 4.79 to 5.64 days. Option $s_t$ and the other two populations change the distances and shares by less than 0.001 and the centroids by less than 0.02 days.

\begin{table}[t]
\centering
\small
\setlength{\tabcolsep}{4pt}
\caption{Organic response of $s_0$ under three intergroup matrices with fixed initial segment responses.}
\label{tab:app-propagation}
\begin{tabular}{@{}lrrr@{}}
\toprule
Matrix & \shortstack{Mean daily TV distance\\from $A$} & \shortstack{Share in targeted\\segments} & \shortstack{Segment time centroids\\(days)} \\
\midrule
Structured $A$ & 0 & 96.05\% & 4.79--5.64 \\
Within-segment & 0.246 & 100\% & 5.02 \\
Uniform cross-segment & 0.067 & 94.26\% & 4.75--5.42 \\
\bottomrule
\end{tabular}
\end{table}

\subsection{Parameter Sensitivity}
\label{app:case-sensitivity}

Each setting changes one value of Table~\ref{tab:app-parameters} and keeps the others: the decay rate $\beta$, the branching parameter $r$, the targeting strength, the RedNote exploration strength $\delta_p$, or both response weight vectors $\mathbf w_{\mathrm c}$ and $\mathbf w_{\mathrm e}$, scaled by a common factor. Table~\ref{tab:app-sensitivity} gives the budget-doubling ratios. The rankings of the five options are identical in all eleven settings: $s_k,s_b,s_0,s_{cb},s_c$ by the collect-first rule; $s_{cb},s_b,s_0,s_k,s_c$ by $\overline M_{14}$; and $s_0,s_k,s_c,s_{cb},s_b$ by response per CNY. Changing $\beta$ or $r$ multiplies every option's total by the same factor, because all options share the paid schedule and the aggregate recursion~\eqref{eq:app-total} is linear, so the ratios stay unchanged. The collect-first ranking uses the fixed note predictions. The settings are a local stress test around the design values.

\begin{table}[t]
\centering
\small
\setlength{\tabcolsep}{5pt}
\caption{Ratio of $\overline M_{14}$ to the baseline $s_0$ for the two CNY~80,000 options under single-parameter changes.}
\label{tab:app-sensitivity}
\begin{tabular}{@{}llrr@{}}
\toprule
Parameter & Values & $s_b/s_0$ & $s_{cb}/s_0$ \\
\midrule
Design values & Table~\ref{tab:app-parameters} & 1.962 & 1.965 \\
Decay rate $\beta$ & 0.6; 1.2 & 1.962; 1.962 & 1.965; 1.965 \\
Branching parameter $r$ & 0.2; 0.5 & 1.962; 1.962 & 1.965; 1.965 \\
Targeting strength & 1.5; 2.5 & 1.950; 1.972 & 1.953; 1.974 \\
Exploration strength $\delta_p$ & 0.325; 0.975 & 1.963; 1.962 & 1.966; 1.965 \\
Response weight scale & 0.8; 1.2 & 1.953; 1.973 & 1.956; 1.975 \\
\bottomrule
\end{tabular}
\end{table}

\FloatBarrier
\section{RedNote Engagement Prediction}
\label{app:prediction}

\subsection{Corpora and Features}
\label{app:prediction-data}

The two corpora support different comparisons. The first contains 39,000 note snapshots and compares the configurations without note age (v3) and with note age (v3.1-pg) on the same five folds, fixed after shuffling the notes. The second contains 12,154 notes from 13 niches, each with at least 100 reads, and is used for the structured holdout experiments.

For the 39,000-note comparison, titles and bodies are embedded separately with OpenAI text-embedding-3-small. Each 1,536-dimensional output keeps its first 768 dimensions and is renormalized to unit length; the two vectors are concatenated into 1,536 dimensions and reduced to 128 principal components. The resulting predictor inputs contain 182 features for v3 and 183 for v3.1-pg. Each outcome regressor uses up to 500 LightGBM trees, learning rate 0.05, maximum depth 6, 32 leaves, a minimum of 20 examples per leaf, and L1 and L2 penalties of 0.3. Five-fold cross-validation uses shared shuffled folds with seed 42 and early stopping after 30 rounds; the regressor seed is 0.

Each structured holdout split refits preprocessing and predictors. Titles and bodies each receive a 768-dimensional deterministic hashed representation; the two are concatenated and reduced to at most 64 dimensions by randomized principal component analysis fitted on the training set only. The remaining features are creator follower count, content duration, media type, niche, the 30 most frequent topics in the training set, cyclic encodings of publication hour and weekday, and note age. Each outcome has its own LightGBM regressor trained on $\log(1+y)$, and $R^2_{\log}$ and RMSE$_{\log}$ are computed on the log scale. The structured holdout regressors use 300 trees, learning rate 0.05, maximum depth 6, 32 leaves, a minimum of 20 examples per leaf, and L1 and L2 penalties of 0.3.

\subsection{Holdout Splits}
\label{app:prediction-splits}

Table~\ref{tab:app-holdout} lists the three holdout splits. The temporal split trains on the earliest 85\% of notes by publication time and tests on the latest 15\%. The unseen-creator split groups notes by creator; none of the 1,642 creators in the test set appears in training. The leave-one-niche-out split holds out each niche in turn, and the metrics are computed on the pooled test predictions of the 13 folds. Table~\ref{tab:app-niches} lists the number of notes in each niche.

\begin{table}[t]
\centering
\small
\setlength{\tabcolsep}{4pt}
\caption{Structured holdout experiments on the 12,154-note corpus.}
\label{tab:app-holdout}
\begin{tabular}{@{}llcrrrr@{}}
\toprule
Split & Train / test & Metric & Reads & Likes & Collects & Comments \\
\midrule
Latest-time & 10,331 / 1,823 & $R^2_{\log}$ & 0.253 & 0.361 & 0.329 & $-0.007$ \\
 & & RMSE$_{\log}$ & 1.879 & 1.972 & 1.831 & 1.525 \\
\addlinespace
Unseen creator & 10,310 / 1,844 & $R^2_{\log}$ & 0.559 & 0.550 & 0.513 & 0.347 \\
 & & RMSE$_{\log}$ & 1.506 & 1.736 & 1.688 & 1.463 \\
\addlinespace
Leave-one-niche-out & 12,154 / 13 folds & $R^2_{\log}$ & 0.503 & 0.513 & 0.456 & 0.333 \\
 & & RMSE$_{\log}$ & 1.625 & 1.811 & 1.781 & 1.500 \\
\bottomrule
\end{tabular}
\end{table}

\begin{table}[t]
\centering
\small
\caption{Niche composition of the holdout corpus.}
\label{tab:app-niches}
\begin{tabular}{@{}lrlrlr@{}}
\toprule
Niche & Notes & Niche & Notes & Niche & Notes \\
\midrule
Fashion & 1,466 & Home & 1,202 & Automotive & 819 \\
Parenting & 1,374 & Pets & 1,062 & Fitness & 618 \\
Food & 1,309 & Electronics & 987 & Health care & 437 \\
Skincare & 1,221 & Education & 970 & Travel & 410 \\
Makeup & 279 & & & & \\
\bottomrule
\end{tabular}
\end{table}

\subsection{Count-Scale Errors}
\label{app:prediction-counts}

Table~\ref{tab:app-counts} reports the raw-count mean absolute error (MAE) and the ratio of the mean predicted count to the mean observed count. For the structured holdouts, both come from the stored MAE and ten-bin calibration aggregates, and mean predictions are 18\%--43\% of the observed means in all three splits and four outcomes. For the 39,000-note predictor used in the campaign case, repeating the five-fold protocol of Table~\ref{tab:prediction} on a re-extraction of the corpus reproduces its $R^2_{\log}$ to within 0.003; out-of-fold mean predictions are 50\%--59\% of the observed means, and their medians are 98\%--107\% of the observed medians. The regressors fit $\log(1+y)$ and report $\exp(\hat z)-1$, which lies below the conditional mean of a right-skewed count.

\begin{table}[t]
\centering
\small
\setlength{\tabcolsep}{4pt}
\caption{Count-scale errors of the RedNote predictors. Ratio is the mean predicted count divided by the mean observed count.}
\label{tab:app-counts}
\begin{tabular}{@{}lrrrrrrrr@{}}
\toprule
 & \multicolumn{2}{c}{Reads} & \multicolumn{2}{c}{Likes} & \multicolumn{2}{c}{Collects} & \multicolumn{2}{c}{Comments} \\
\cmidrule(lr){2-3}\cmidrule(lr){4-5}\cmidrule(lr){6-7}\cmidrule(l){8-9}
Split & MAE & Ratio & MAE & Ratio & MAE & Ratio & MAE & Ratio \\
\midrule
39,000-note five-fold CV & 62,594 & 0.59 & 3,116 & 0.54 & 1,145 & 0.52 & 226.6 & 0.50 \\
\midrule
Latest-time & 17,305 & 0.24 & 1,025 & 0.18 & 332 & 0.21 & 42.7 & 0.39 \\
Unseen creator & 43,423 & 0.43 & 2,232 & 0.36 & 876 & 0.31 & 145.0 & 0.37 \\
Leave-one-niche-out & 50,018 & 0.40 & 2,329 & 0.34 & 848 & 0.30 & 178.9 & 0.31 \\
\bottomrule
\end{tabular}
\end{table}

\subsection{Note-Age Reference}
\label{app:prediction-age}

In training, note age counts the days from publication to the date the predictor is fitted, capped at 180 days, whereas the engagement counts were collected earlier. To measure how this reference affects temporal evaluation, a re-extracted sample of 12,154 notes with the size and niche composition of the holdout corpus is split by whole publication dates into 9,265 training notes (March 17--26, 2026), 1,749 calibration notes (March 27), and 1,140 test notes (March 28 to April 19). The recorded update dates of these rows range from March 24 to April 26. Hashed-text LightGBM regressors are refitted with note age counted either to the extraction date, September 25, 2026, which caps 96.8\% of ages at 180 days, or to the date of the original holdout run, May 5, 2026, which caps none.

Changing only this reference lowers test MAE by 43\%--58\% across the four outcomes (Table~\ref{tab:app-age}). Under both references the test $R^2_{\log}$ is negative; this split trains on ten publication days and tests on the following three weeks. A log-linear recalibration fitted on the calibration day raises $R^2_{\log}$ under the May reference to 0.158, 0.258, 0.227, and $-0.068$, while mean predictions remain 70\%--85\% below the observed means. A multiplicative recalibration that matches the calibration-day mean count increases MAE for all four outcomes.

\begin{table}[t]
\centering
\small
\setlength{\tabcolsep}{4pt}
\caption{Temporal diagnostic on a re-extracted 12,154-note sample under two note-age references. MAE is on the count scale; the last two columns use the May reference with log-linear recalibration.}
\label{tab:app-age}
\begin{tabular}{@{}lrrrrr@{}}
\toprule
 & \multicolumn{2}{c}{Test MAE} & & \multicolumn{2}{c}{Log-linear recalibration} \\
\cmidrule(lr){2-3}\cmidrule(l){5-6}
Outcome & September reference & May reference & $R^2_{\log}$, May & MAE & Mean bias \\
\midrule
Reads & 29,306 & 13,446 & $-0.508$ & 11,332 & $-82.0$\% \\
Likes & 1,182 & 676 & $-0.260$ & 586 & $-84.3$\% \\
Collects & 507 & 291 & $-0.215$ & 264 & $-85.3$\% \\
Comments & 103.3 & 43.8 & $-1.001$ & 31.6 & $-70.1$\% \\
\bottomrule
\end{tabular}
\end{table}

\FloatBarrier
\section{Policy Value and Audience Ranking}
\label{app:public}

\subsection{Off-Policy Estimators and Identification}
\label{app:ope}

All four estimators of Section~\ref{subsec:policy-evaluation} use the importance weight $w_i=\pi(a_i\mid x_i)/\mu_i$. DR and Switch-DR \citep{dudik2011doubly,wang2017optimal} also use an outcome model $\hat q(a)$ that shrinks the mean reward of each action in the earlier part of a log toward that part's mean reward,
\begin{equation}
\hat q(a)=\frac{\sum_{j:a_j=a}y_j+\beta_q\bar y}{|\{j:a_j=a\}|+\beta_q},\qquad \beta_q=20,
\end{equation}
where the sums run over the earlier part and $\bar y$ is its mean reward; on Open Bandit, $\hat q$ is indexed by action and recommendation position. The model value of the target policy in context $x$ is $V_\pi(x)=\sum_a\pi(a\mid x)\hat q(a)$. With residual $\delta_i=y_i-\hat q(a_i)$,
\begin{equation}
\hat V_{\mathrm{DR}}(\pi)=\frac1n\sum_{i=1}^n\bigl[V_\pi(x_i)+w_i\delta_i\bigr],\qquad
\hat V_{\mathrm{Switch\text{-}DR}}(\pi)=\frac1n\sum_{i=1}^n\bigl[V_\pi(x_i)+w_i\,\mathbb 1\{w_i\le\tau\}\,\delta_i\bigr]
\end{equation}
Switch-DR applies the residual correction only to records whose weight does not exceed $\tau$. The KuaiRand-Pure experiment uses $\tau=100$ and the Open Bandit experiment $\tau=5$, with the logged propensity scores and recommendation positions of the Open Bandit Dataset \citep{saito2020open}. Because the outcome model does not depend on user context, the model value is the same in every context.

Importance weighting identifies the policy value under three conditions: (1) overlap, $\pi_0(a\mid x)>0$ for every context $x$ and every action $a$ with $\pi(a\mid x)>0$; (2) unconfoundedness, potential outcomes are independent of the logged action given the context; and (3) the stable unit treatment value assumption, under which a unit's potential outcome depends only on its own action. Under these conditions,
\begin{equation}
V(\pi)=\mathbb E_{x,\,a\sim\pi}[Y(a)]
=\mathbb E_x\Bigl[\sum_a\pi(a\mid x)\,\mathbb E[Y\mid x,a]\Bigr]
=\mathbb E_{x,\,a\sim\pi_0}\Bigl[\frac{\pi(a\mid x)}{\pi_0(a\mid x)}\,Y\Bigr],
\end{equation}
where the second equality uses unconfoundedness and the third uses overlap \citep{li2011unbiased}. KuaiRand-Pure inserts videos sampled uniformly from its candidate pool, and the Open Bandit Dataset records the assignment probability of every logged record.

The target policy and outcome model come from an earlier part of each log, and the estimates use the later part. KuaiRand-Pure splits its random log by calendar date. April 22--29, 2022, with 337,933 records, selects the ten most-logged videos among the 7,583 candidates, breaking ties by identifier, and fits $\hat q$; April 30 to May 8 is evaluated after removing 153 records whose timestamps precede the last construction event, which leaves 847,973. Open Bandit splits each log at its median timestamp. The two earlier halves together select one set of ten actions, each earlier half fits its own $\hat q$, and each later half is evaluated after the last construction timestamp of both logs, leaving 5,000 random and 4,964 BTS records. A selected action has target probability $0.1/80+0.9/10$ at every recommendation position, and any other action $0.1/80$. Intervals are percentile intervals from 1,000 bootstrap resamples of users on KuaiRand and of equal-timestamp groups on Open Bandit, conditional on the fixed target policy and outcome model.

When the weights use the probabilities with which the log was generated, the mean importance weight has expectation one. It is 1.060 with interval $[1.005,1.114]$ on KuaiRand, 0.904 with $[0.841,0.969]$ on the random Open Bandit log, and 0.969 with $[0.903,1.045]$ on the BTS log. The KuaiRand logging probability $1/7{,}583$ follows the uniform sampling of randomly inserted videos described with the dataset \citep{gao2022kuairand}. The candidate pool of the random log changes over time. Treating each video as a candidate from the first to the last day on which it is logged, the pool shrinks from 7,475 videos on April 23 to 7,214 on May 8: 369 videos stop appearing before May 8, and 72 first appear after April 23. Per-video exposure counts over the whole log have a dispersion index of 6.08, against 1.00 under uniform sampling from the 7,583 videos; among the 7,198 videos logged by April 23 and still logged on May 8, the index is 1.17, and the correlation between their construction and evaluation counts is 0.05. If each day's videos are drawn uniformly from that day's pool, weights computed with $1/7{,}583$ have an expected mean of 1.036 for the target policy. The ten videos it selects are in the pool on every evaluation day, and the daily pools cover 99.61\% of its probability on average over the evaluated records. Replacing $1/7{,}583$ by the reciprocal of each day's pool size lowers the observed mean weight to 1.019 and gives SNIPS 0.0880, compared with 0.0879 under $1/7{,}583$. Building the KuaiRand target policy from the standard log of April 8--21 instead, and evaluating it on the whole random log, gives SNIPS 0.2545 with interval $[0.2303,0.2762]$, an effective sample size of 1,596, and a mean weight of 0.804 with interval $[0.767,0.845]$. This target policy differs from the one in Table~\ref{tab:public}, so the two values describe different policies.

\begin{table}[t]
\centering
\small
\setlength{\tabcolsep}{3pt}
\caption{Off-policy estimates on the later part of each log. ESS, mean importance weight, and weight range describe how each log supports the target policy; the mean weight has expectation one when the weights use the probabilities with which the log was generated.}
\label{tab:public}
\begin{tabularx}{\linewidth}{@{}>{\raggedright\arraybackslash}Xrrrrrrl@{}}
\toprule
Data / logging policy & IPS & SNIPS & DR & Switch-DR & ESS & Mean weight & Weight range \\
\midrule
KuaiRand / uniform & 0.0932 & 0.0879 & 0.0883 & 0.0828 & 1,454 & 1.060 & $[0.10,682.57]$ \\
Open Bandit / random & 0.00902 & 0.00998 & 0.00915 & 0.00280 & 685 & 0.904 & $[0.10,7.30]$ \\
Open Bandit / BTS & 0.00209 & 0.00216 & 0.00258 & 0.00396 & 611 & 0.969 & $[0.0027,109.94]$ \\
\bottomrule
\end{tabularx}
\end{table}

\subsection{Uplift Metrics}
\label{app:uplift}

Test customers are sorted by predicted effect $\hat\tau(x_i)$ in decreasing order. Let $T_m$ be the first $m$ customers, with treatment and control members $T_m^{(1)}$ and $T_m^{(0)}$. The gain curve is
\begin{equation}
U(m)=\bigl[\bar y(T_m^{(1)})-\bar y(T_m^{(0)})\bigr]\,m .
\end{equation}
Let $A(\hat\tau)$ be the area under the gain curve, $A_{\mathrm{rand}}$ the area under the line from $(0,0)$ to $(n,U(n))$, and $A_{\mathrm{perfect}}$ the area under the curve of the best ordering constructed from observed outcomes. The normalized AUUC is $(A(\hat\tau)-A_{\mathrm{rand}})/(A_{\mathrm{perfect}}-A_{\mathrm{rand}})$, which is 0 for a random ordering and 1 for the best ordering. The Qini metric replaces $U(m)$ with
\begin{equation}
Q(m)=\sum_{i\in T_m}y_i\,\mathbb 1\{a_i=1\}-\Bigl(\sum_{i\in T_m}y_i\,\mathbb 1\{a_i=0\}\Bigr)\frac{|T_m^{(1)}|}{|T_m^{(0)}|}
\end{equation}
and is normalized in the same way. Uplift at a reach fraction is defined in Section~\ref{subsec:ranking-evaluation}. The three metrics match \texttt{uplift\_auc\_score}, \texttt{qini\_auc\_score}, and \texttt{uplift\_at\_k} with \texttt{strategy="overall"} in scikit-uplift \citep{radcliffe2011quality,renaudin2021about}.

\subsection{Case Study: Audience Ranking under Limited Reach}
\label{app:x5}

\begin{figure}[t]
\centering
\includegraphics[width=0.78\linewidth]{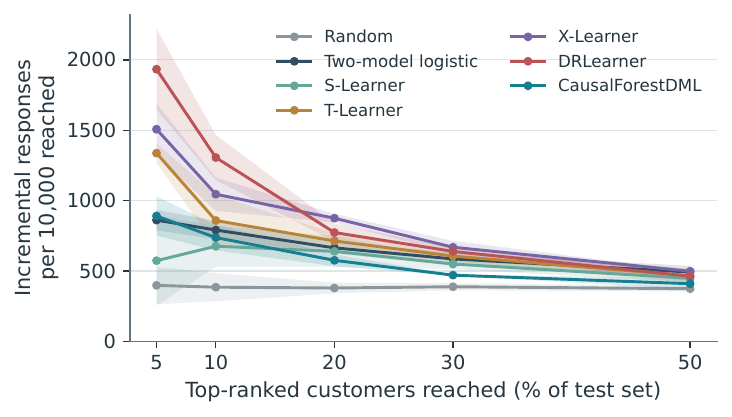}
\caption{Estimated incremental responses per 10,000 customers reached on X5 RetailHero, by the fraction of top-ranked customers reached. Lines are means over three seeds and bands show $\pm1$ standard deviation.}
\label{fig:app-x5}
\end{figure}

In X5 RetailHero, the advertiser sends a marketing message to randomly selected customers, with a treatment fraction of 0.4998. A campaign often reaches only part of its customers, so this case compares which ranking method yields the largest incremental response at each reach fraction. For each seed in $\{42,137,256\}$, the data are split 75\%/25\% into training and test sets, stratified jointly by treatment and outcome; all seven methods share the training and test customers, and preprocessing is fitted on the training set only. S-Learner, T-Learner, X-Learner, and DRLearner use 80-tree forests as base learners, CausalForestDML uses 200 trees, and two-model logistic regression fits separate logistic regressions to the treatment and control groups. All forests use a minimum leaf size of 50. The two logistic regressions and the logistic propensity model use the lbfgs solver with at most 500 iterations. DRLearner uses three-fold cross-fitting with separate 80-tree regression and final forests. CausalForestDML uses 80-tree random-forest classifiers for its outcome and treatment nuisance models, three-fold cross-fitting, and 200 causal trees. Table~\ref{tab:app-x5} reports all methods, and Figure~\ref{fig:app-x5} converts uplift into estimated incremental responses per 10,000 customers reached. DRLearner is highest when the campaign reaches the top 5\% or 10\% of customers, and X-Learner is highest from 20\% to 50\%; as the reach fraction grows, the incremental responses of all methods move closer to random ranking.

\begin{table*}[t]
\centering
\small
\setlength{\tabcolsep}{3.5pt}
\caption{Audience ranking on X5 RetailHero. Entries are means $\pm$ standard deviations over three seeds; uplift columns are headed by the reach fraction. The highest mean in each column is in bold.}
\label{tab:app-x5}
\textit{(a) Aggregate ranking metrics}\par\smallskip
\begin{tabular*}{\linewidth}{@{\extracolsep{\fill}}lcc@{}}
\toprule
Method & AUUC & Qini \\
\midrule
Random & $0.0040\pm0.0016$ & $0.0025\pm0.0009$ \\
Two-model logistic & $\mathbf{0.0217\pm0.0019}$ & $0.0140\pm0.0013$ \\
S-Learner & $0.0125\pm0.0038$ & $0.0082\pm0.0025$ \\
T-Learner & $0.0155\pm0.0054$ & $0.0105\pm0.0036$ \\
X-Learner & $0.0213\pm0.0030$ & $\mathbf{0.0145\pm0.0021}$ \\
DRLearner & $0.0202\pm0.0005$ & $0.0141\pm0.0004$ \\
CausalForestDML & $0.0112\pm0.0048$ & $0.0074\pm0.0033$ \\
\bottomrule
\end{tabular*}
\par\medskip
\textit{(b) Uplift at 5\%, 10\%, and 20\% reach}\par\smallskip
\begin{tabular*}{\linewidth}{@{\extracolsep{\fill}}lccc@{}}
\toprule
Method & Uplift@5\% & Uplift@10\% & Uplift@20\% \\
\midrule
Random & $0.0397\pm0.0133$ & $0.0385\pm0.0100$ & $0.0379\pm0.0038$ \\
Two-model logistic & $0.0861\pm0.0073$ & $0.0791\pm0.0066$ & $0.0664\pm0.0041$ \\
S-Learner & $0.0573\pm0.0320$ & $0.0676\pm0.0144$ & $0.0639\pm0.0110$ \\
T-Learner & $0.1337\pm0.0069$ & $0.0858\pm0.0182$ & $0.0713\pm0.0053$ \\
X-Learner & $0.1507\pm0.0182$ & $0.1045\pm0.0118$ & $\mathbf{0.0875\pm0.0030}$ \\
DRLearner & $\mathbf{0.1933\pm0.0287}$ & $\mathbf{0.1306\pm0.0160}$ & $0.0773\pm0.0064$ \\
CausalForestDML & $0.0890\pm0.0139$ & $0.0737\pm0.0091$ & $0.0576\pm0.0042$ \\
\bottomrule
\end{tabular*}
\par\medskip
\textit{(c) Uplift at 30\% and 50\% reach}\par\smallskip
\begin{tabular*}{\linewidth}{@{\extracolsep{\fill}}lcc@{}}
\toprule
Method & Uplift@30\% & Uplift@50\% \\
\midrule
Random & $0.0388\pm0.0025$ & $0.0374\pm0.0019$ \\
Two-model logistic & $0.0585\pm0.0045$ & $0.0492\pm0.0043$ \\
S-Learner & $0.0549\pm0.0037$ & $0.0448\pm0.0029$ \\
T-Learner & $0.0606\pm0.0038$ & $0.0465\pm0.0050$ \\
X-Learner & $\mathbf{0.0668\pm0.0046}$ & $\mathbf{0.0498\pm0.0002}$ \\
DRLearner & $0.0639\pm0.0026$ & $0.0462\pm0.0034$ \\
CausalForestDML & $0.0470\pm0.0004$ & $0.0409\pm0.0038$ \\
\bottomrule
\end{tabular*}
\end{table*}

\FloatBarrier
\section{Paired Counterfactual Scoring}
\label{app:paired}

\subsection{Scenario Generator and Splits}
\label{app:paired-data}

The scenarios come from the synthetic scenario generator in the OranSim open-source repository at commit \texttt{8b5dd54}, run with seed 2027 to produce 20,000 scenarios. The treatment arm of each scenario is determined jointly by its budget tier and creator tier, $\text{arm}=\min\bigl(3,\lfloor\text{budget tier}/2\rfloor+2\cdot\mathbb 1[\text{creator tier is mid or above}]\bigr)$. The alternative setting moves the scenario to arm $(\text{arm}+1)\bmod 4$ and multiplies the budget by that arm's factor (0.6, 1.0, 1.4, and 1.8 for arms 0--3), keeping the other creator attributes fixed. The generator computes exposure, clicks, conversions, and revenue under both settings. In lexical order of scenario identifiers, the first 16,000 scenarios form the training set and the next two blocks of 2,000 form the validation and test sets, with no overlap. An identity intervention that keeps the arm and budget unchanged reuses the same computed potential outcome, so its effect is exactly zero.

The feature-substitution baseline uses platform, niche, budget tier, creator tier, treatment arm, log budget, log follower count, and creator engagement rate. Each outcome has its own LightGBM regressor (300 trees, learning rate 0.05, maximum depth 6, 32 leaves, at least 20 samples per leaf, and L1 and L2 regularization of 0.3) trained on $\log(1+y)$.

\subsection{Full Metrics and Segment-Level Effects}
\label{app:paired-results}

\begin{figure}[t]
\centering
\includegraphics[width=\linewidth]{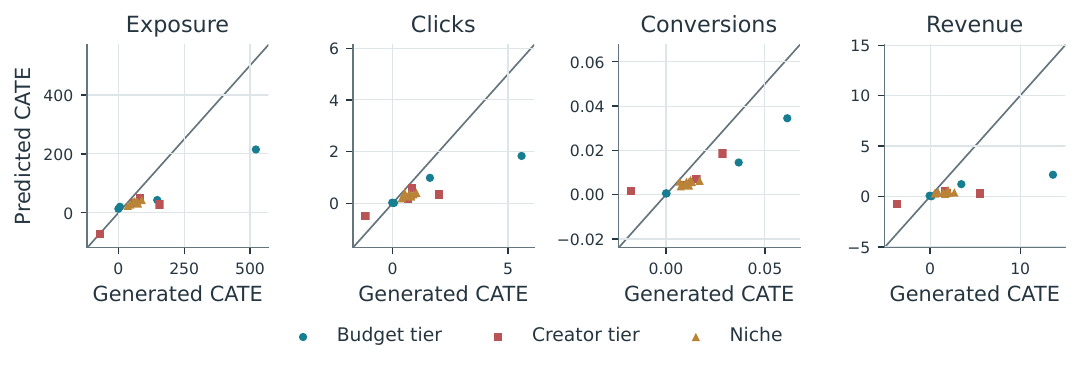}
\caption{Grouped CATE of the LightGBM feature-substitution baseline. Each point is a budget tier, creator tier, or niche; the horizontal axis is the within-group mean of generated effects and the vertical axis the within-group mean of predicted effects. The diagonal marks equality.}
\label{fig:app-paired}
\end{figure}

Table~\ref{tab:app-paired} extends Table~\ref{tab:paired} with the mean absolute errors of factual outcomes, counterfactual outcomes, and individual effects, and with the rank correlation of individual effects. The grouped CATE $R^2$ is computed over 18 groups: four budget tiers, four creator tiers, and ten niches. Figure~\ref{fig:app-paired} plots the generated and predicted effects of these 18 groups.

\begin{table*}[t]
\centering
\small
\setlength{\tabcolsep}{4pt}
\caption{All metrics of the LightGBM feature-substitution baseline on the 2,000 paired synthetic test scenarios. Counterfactual normalized MAE divides the counterfactual MAE by the mean absolute alternative outcome.}
\label{tab:app-paired}
\textit{(a) Factual and counterfactual outcome prediction}\par\smallskip
\begin{tabular*}{\linewidth}{@{\extracolsep{\fill}}lcccc@{}}
\toprule
Outcome & \shortstack{Factual\\$R^2$} & \shortstack{Factual\\MAE} & \shortstack{Counterfactual\\MAE} & \shortstack{Counterfactual\\normalized MAE} \\
\midrule
Exposure & 0.964 & 18.24 & 61.14 & 0.219 \\
Clicks & 0.892 & 0.477 & 0.810 & 0.256 \\
Conversions & 0.762 & 0.0055 & 0.0121 & 0.582 \\
Revenue & 0.667 & 1.729 & 2.657 & 0.439 \\
\bottomrule
\end{tabular*}
\par\medskip
\textit{(b) Individual and grouped effect estimation}\par\smallskip
\begin{tabular*}{\linewidth}{@{\extracolsep{\fill}}lcccc@{}}
\toprule
Outcome & \shortstack{Individual-effect\\MAE} & \shortstack{Individual-effect\\sign accuracy} & \shortstack{Individual-effect\\rank correlation} & \shortstack{Grouped\\CATE $R^2$} \\
\midrule
Exposure & 63.46 & 0.955 & 0.544 & 0.457 \\
Clicks & 0.920 & 0.617 & 0.429 & 0.355 \\
Conversions & 0.0138 & 0.930 & 0.450 & 0.531 \\
Revenue & 3.230 & 0.552 & 0.233 & 0.003 \\
\bottomrule
\end{tabular*}
\end{table*}

\clearpage
\section{Reproducibility Materials}
\label{app:repro}

\subsection{Deterministic Output Snapshot}
\label{app:snapshot}

The output snapshot contains 3,000 scenario scoring records, 1,000 for each of seeds 42, 137, and 256, generated by the v3.1-pg predictor. The scenarios cover 15 niches, three creator tiers (nano, mid, and head), three budget tiers, two exposure regimes (platform recommendation and random), and image and video posts. Each record contains the scenario fields (niche, title, post type, duration, image count, creator tier, follower count, budget tier and amount, exposure regime, and note age), the four predicted engagement counts and their per-follower rates, and the predictor version and seed. Records also carry an exposure field set to eight times the predicted reads, which this paper does not use. Text representations are deterministic hashes; regenerating the snapshot with the same seeds and record count produces byte-identical files. Table~\ref{tab:app-snapshot} summarizes the four predictions.

\begin{table}[htbp]
\centering
\small
\caption{Predicted engagement counts in the output snapshot.}
\label{tab:app-snapshot}
\begin{tabular}{@{}lrr@{}}
\toprule
Outcome & Mean & Standard deviation \\
\midrule
Reads & 119,938.78 & 51,215.30 \\
Likes & 4,605.42 & 2,537.27 \\
Collects & 768.90 & 374.68 \\
Comments & 355.15 & 205.73 \\
\bottomrule
\end{tabular}
\end{table}

\subsection{Released and Private Materials}
\label{app:materials}

The public OranSim repository contains the simulator code, the synthetic scenario generator and its scenarios, and the scripts and aggregate results of the KuaiRand-Pure, Open Bandit, X5, and paired counterfactual experiments. The note predictors are trained on a private RedNote corpus, and Appendices~\ref{app:case} and~\ref{app:prediction} report the aggregate results of the RedNote experiments.

\end{document}